\documentclass[%
 aip,
 amsmath,amssymb,
 reprint,%
]{revtex4-1}

\usepackage{graphicx}
\usepackage{xcolor}
\usepackage{dcolumn}
\usepackage{bm}

\usepackage[utf8]{inputenc}
\usepackage[T1]{fontenc}
\usepackage{mathptmx}
\usepackage{amsmath}

\usepackage{ulem}
\begin{document}

\makeatletter
    \def\@fnsymbol#1{\ensuremath{\ifcase#1\or \dagger\or \ddagger\or
       \mathsection\or \mathparagraph\or \|\or **\or \dagger\dagger
       \or \ddagger\ddagger \else\@ctrerr\fi}}
\makeatother

\preprint{AIP/123-QED}
\title{Exploring the robust extrapolation of high-dimensional machine learning potentials.}
\author{Claudio Zeni}
\email{claudio.zeni@sissa.it}
\affiliation{Physics Area, International School for Advanced Studies,  Trieste, IT}
\author{Andrea Anelli}
\affiliation{Roche Pharma Research and Early Development, Therapeutic Modalities, Roche Innovation Center Basel, F.Hoffmann-La Roche Ltd, Grenzacherstrasse 124, 4070 Basel, Switzerland }
\author{Aldo Glielmo}
\affiliation{Physics Area, International School for Advanced Studies,  Trieste, IT}
\affiliation{Banca d'Italia, Roma,  Italy}
\thanks{The views and opinions expressed in this paper are those of the authors and do not necessarily reflect the official policy or position of Banca d’Italia.}
\author{Kevin Rossi}%
\email{kevin.rossi@epfl.ch} 
\affiliation{Laboratory of Nanochemistry, Institute of Chemistry and Chemical Engineering, Ecole Polytechnique Fédérale de Lausanne, Lausanne, Switzerland}%
%


\begin{abstract}

We show that, contrary to popular assumptions, predictions from machine learning potentials built upon high-dimensional atom-density representations almost exclusively occur in regions of the representation space which lie outside the convex hull defined by the training set points.
We then propose a perspective to rationalize the domain of robust extrapolation and accurate prediction of atomistic machine learning potentials in terms of the probability density induced by training points in the representation space.
\keywords{local density representation, extrapolation, interpolation, machine learning}
\end{abstract}

\maketitle

Machine learning (ML) potentials for atomistic systems infer the mapping between configurations and a target objective function, i.e., the total energy of the system and the forces acting on each atom.
These potentials are trained on a database of configurations whose objective function has been calculated via a computationally expensive, yet accurate, reference, e.g., density functional theory (DFT) methods.
Following the training procedure, ML potentials offer predictions that are accurate with respect to, and much faster to compute than, the reference method.
\cite{Musil2021,Deringer2021,Unke2021,Keith2021,Zeni2019}
A key aspect towards the making of accurate and efficient ML potentials lies in the choice of the representation function, which maps atomic coordinates to a set of numerical features.
Among the most successful ones, we find expansions of local densities around atoms in the systems.
In a nutshell, these representations are built upon the description of an atomic environment in terms of atom-centered distributions (encoding $N$-body correlations up to a desired order of $N$), which are approximated via a truncated expansion in radial and angular basis sets. \cite{Behler:2007fe, Bartok2010,Thompson:2015dw,Shapeev2016,Willatt2019,Drautz2019}

One notable characteristic of these representations is that they originate a high-dimensional feature space.
The accuracy of the ML predictions is moreover generally observed to correlate with the representation's dimensionality, when the other free parameters in the expansion are fixed.
\cite{Musil2021, Deringer2021, zuo2020performance}
For this reason, there is an incentive to employ high-dimensional atomic-density representations when training ML potentials.
Recent publications by \citet{Balestriero2021} and \citet{Yousefzadeh2021}\ , built upon the theoretical results presented by \citet{Barany}, showcased that the predictions made by image recognition models whose inputs are high-dimensional, happen in an {\it extrapolation regime}, where
interpolation and extrapolation regimes are formally defined according to a geometric criterion, in particular: \cite{Balestriero2021}
{\it Interpolation occurs for a point $\mathbf{x^*}$ whenever the latter belongs to the Convex Hull (CH) of a set of training points  {\bf X} $\triangleq$ {\bf x$_1$ }, {\bf x$_2$}, ... {\bf x$_M$}, if not, extrapolation occurs.}

The above definition has been employed also in the community of scientists applying ML methods to atomistic systems, and a common assumption is that the accuracy of ML potentials is strongly dependent on the fact that their predictions take place in an interpolation regime.
When a ML potential accurately predicts the objective function for a structure outside the training database, this result is often interpreted as a sign that the atomic environment representations in the out-of-sample structures are "contained in", "covered by", or "interpolated between" points in the training set.
\cite{Bartok2018a, Rossi2020, Tkatchenko2021, Monserrat2020, Nguyen2018, Bartok2018, Zeni2021,  Engel2019, Shao2021}

\
Atom-density representations are built so to naturally encode physical symmetries \cite{Behler:2007fe, Bartok2010,Thompson:2015dw,Shapeev2016,Willatt2019,Drautz2019}.
Further, protocols underlying the generation of atomistic configuration databases hinge on the sampling of trajectories and of physically relevant phases.
These properties reduce the effective degrees of freedom possessed by such representations when compared, e.g., to images.
It is therefore not trivial to predict a priori whether ML potentials exploiting high-dimensional atom-density representations also suffer from a curse of dimensionality, and whether they carry on predictions in an {\it extrapolative regime}, defined according to the previously mentioned geometric criterion.
First, we show that, for the example case of datasets and benchmarks widely quoted in the literature, ML potentials predictions generally occur at point outside the training set convex-hull.
We then propose a protocol to measure the sampling density induced by the training data on points belonging to the test data set.
This quantity is computed on a test point's features as the log probability density of training points estimated via an adaptive k-nearest-neighbors algorithm.
We show that such measure strongly correlates with the prediction errors incurred by ridge-regression potentials on test sets, thus providing an effective tool to identify low-accuracy regions in the representation space, and to rationalize the accuracy of ML potentials according to a rigorous geometric criterion.
We thus clarify the difference between the convex-hull-derived definition of interpolation, and the alternative concept of a well-sampled region in the representation space.
While the knowledge of whether a test point lies within the training set convex-hull yields no information on the test accuracy of a high-dimensional ML potential, we show that one can establish a relationship between the test error incurred by ML potentials and the probability density function induced by the training set and computed on a test point.

To draw general conclusions, we consider three datasets in our investigation.
These comprise periodic and finite-size systems with different chemistries:
\begin{itemize}
    \item The ice-water dataset by \citet{Monserrat2020}, which was employed to test the transferability of a ML force-field trained on water configurations to the case of ice crystals. 
It encompasses a training set of forces and energies in 1000 liquid water configurations -- corresponding to 192000 atomic environments -- and a test set containing structures corresponding to 54 known ice phases -- comprising 2847 atomic environments -- which also includes all the experimentally verified ice structures.
\item The Li, Mo, Ge, Si, Ni, Cu dataset by \citet{zuo2020performance}, which was used to benchmark cost and accuracy of several ML force-field flavors.
It gathers energies and forces in systems of the six different elements for their ground state crystalline bulk configuration, strained crystals, low Miller index surfaces, bulk structures sampled during ab initio molecular dynamics (MD) trajectories at different temperatures, and bulk structures with a vacancy, also sampled during ab initio MD trajectories.
Configurations are then organized, according to a random 90:10 split, into a training and testing set.
\item The Au$_{13}$ database, which was custom-built to probe the likelihood of extrapolation during a MD trajectory.
It comprises five subsets of 1000 configurations of planar Au$_{13}$ nanocluster, with energies and forces labels, sampled every 3~ps during finite-temperature (50K, 100K, 200K, 300K, and 400K) MD runs where no structural rearrangements were observed; for further details we refer the reader to section C of the Supplemental Material.
\end{itemize}

To associate features to atomic environments, we employ the Atom-Centered Symmetry Functions (ACSF),\cite{Behler:2007fe, Singraber2019} the Smooth Overlap of Atomic Positions (SOAP),\cite{Bartok2010, dscribe} or the Atomic Cluster Expansion (ACE) \cite{Drautz2019, Dusson2022} representation (see Supplemental Material, sections A and B, and Tables S1-S6 for further detail), and adopt previously-reported set-ups, when these are available in the literature.\cite{Monserrat2020, Cheng2019, zuo2020performance, Zeni2021descriptors}
We then transform the high-dimensional representations in a linear fashion via principal components analysis (PCA), and construct sets of $P$-dimensional representations which employ the first $P$ principal components, using the same procedure as reported in \citet{Zeni2021descriptors}. 
We then systematically investigate whether  test points are contained within the CH of the training set, also as a function of the number of employed PCA components.
This approach parallels emerging protocols in the literature where low-dimensional embedding of atom-density representations are employed as a tool to probe the similarity among structures \cite{Engel2018, Anelli2018}
and rationalize ML accuracy and transferability. \cite{Musil2021, zuo2020performance, Monserrat2020, Nguyen2018, Bartok2018, Zeni2021,  Engel2019}
Rather than computing the CH -- a time-consuming task in high dimensions \cite{Nielsen1996} -- we 
verify whether a test point $\mathbf{x^*}$ can be expressed as a linear combination of the points in the training set $\{ \mathbf{x}_i \}_{i=1}^M$ constrained to non-negative coefficients $\lambda_i$ summing up to one:
\begin{align}
& \mathbf{x^*}  = \sum_{i=1}^{M} \lambda_i \mathbf{x}_i & \\ \nonumber
\text{with}  ~~~~~ & \sum_{i=1}^{M} \lambda_i = 1 ~~~ \land ~~~\lambda_i \geq 0 ~ \forall ~ i=1,..,M . 
\end{align}
The test point $\mathbf{x^*}$ is in the CH of the training set if and only if the above can be satisfied; this can be verified efficiently via a linear programming approach. \cite{Pardalos1995,Sierksma2015}

\begin{figure}[t!]
    \centering
    \includegraphics[width=8.25cm]{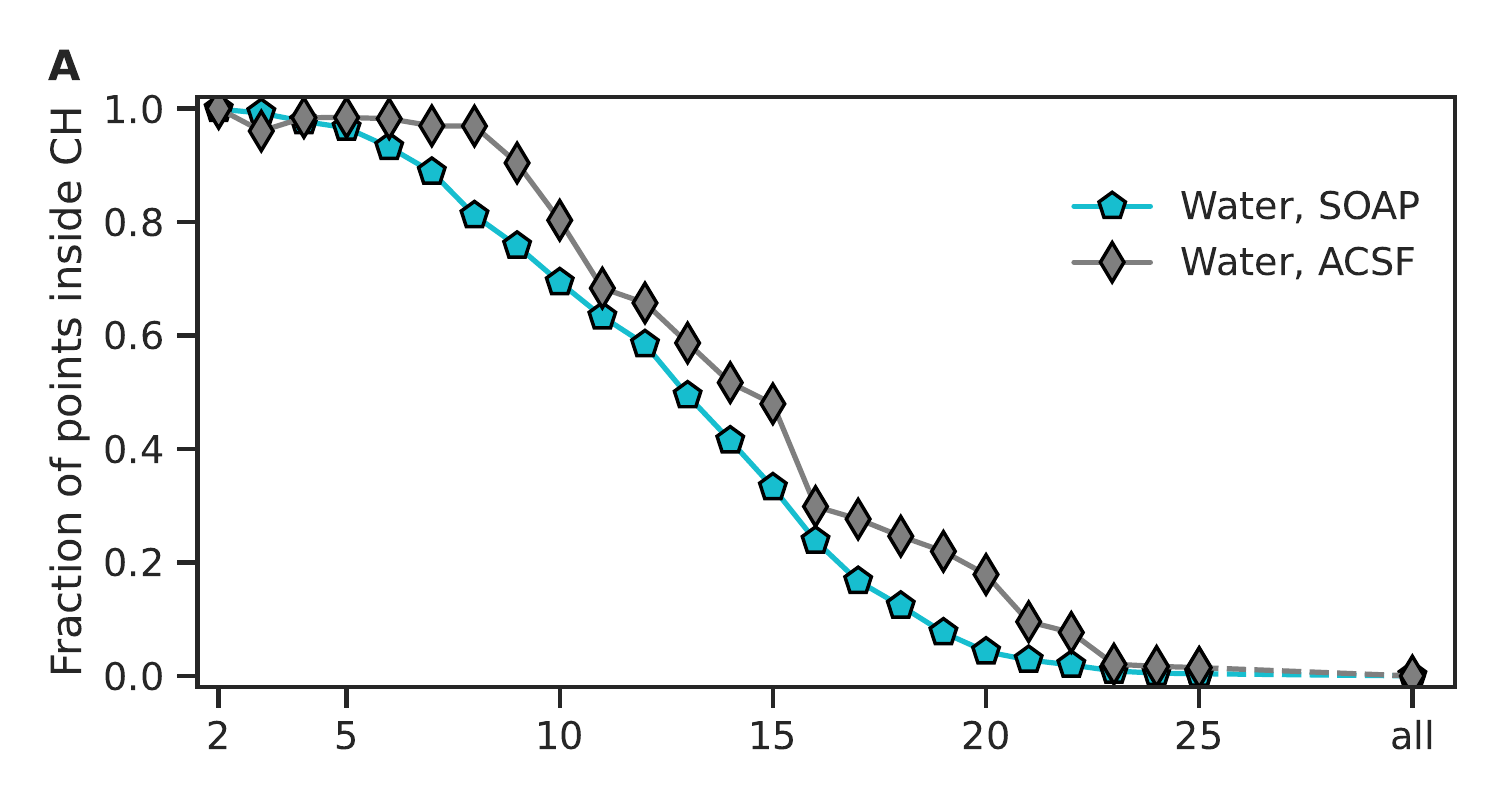}
    \includegraphics[width=8.25cm]{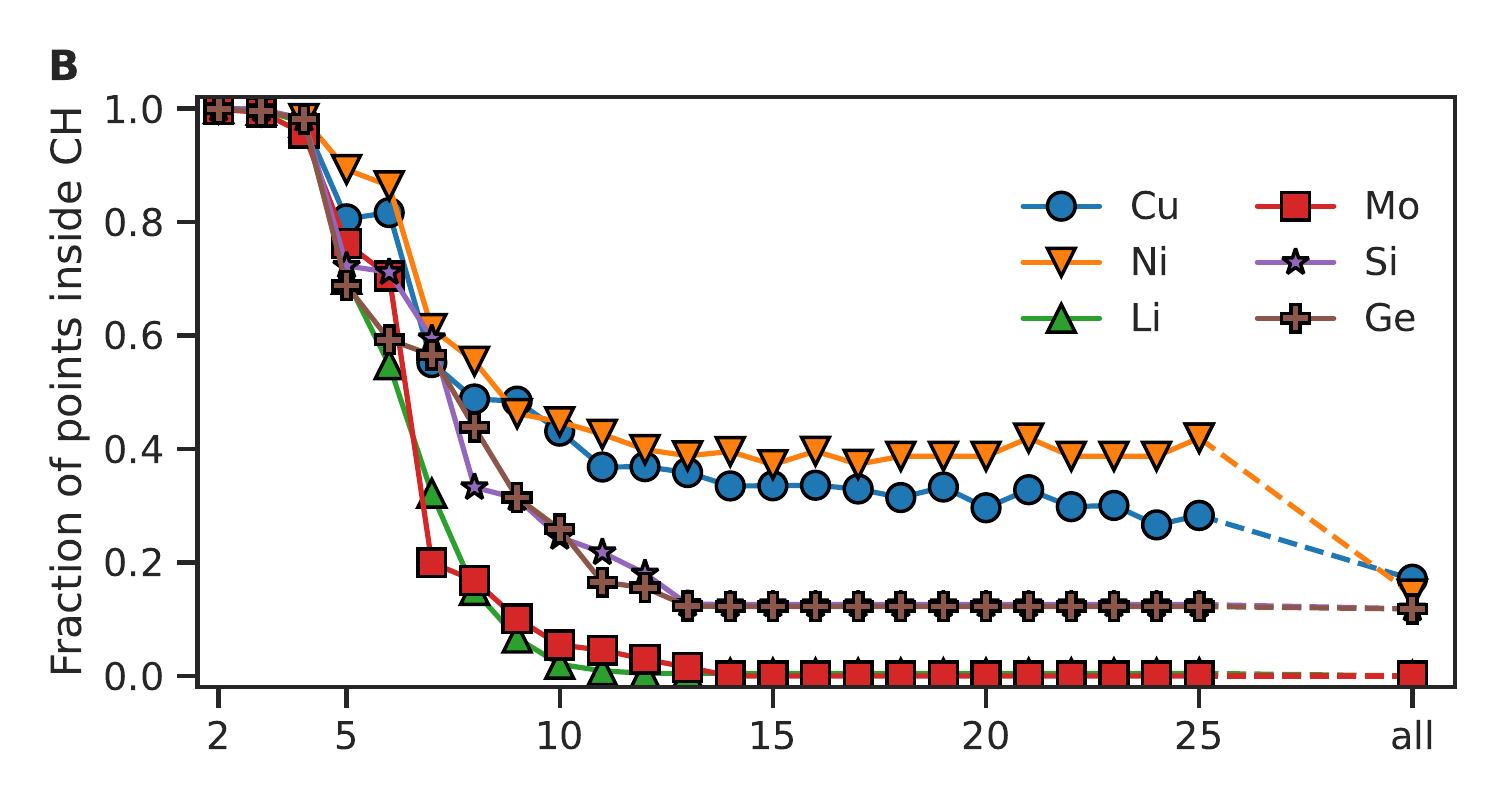}
   \includegraphics[width=8.25cm]{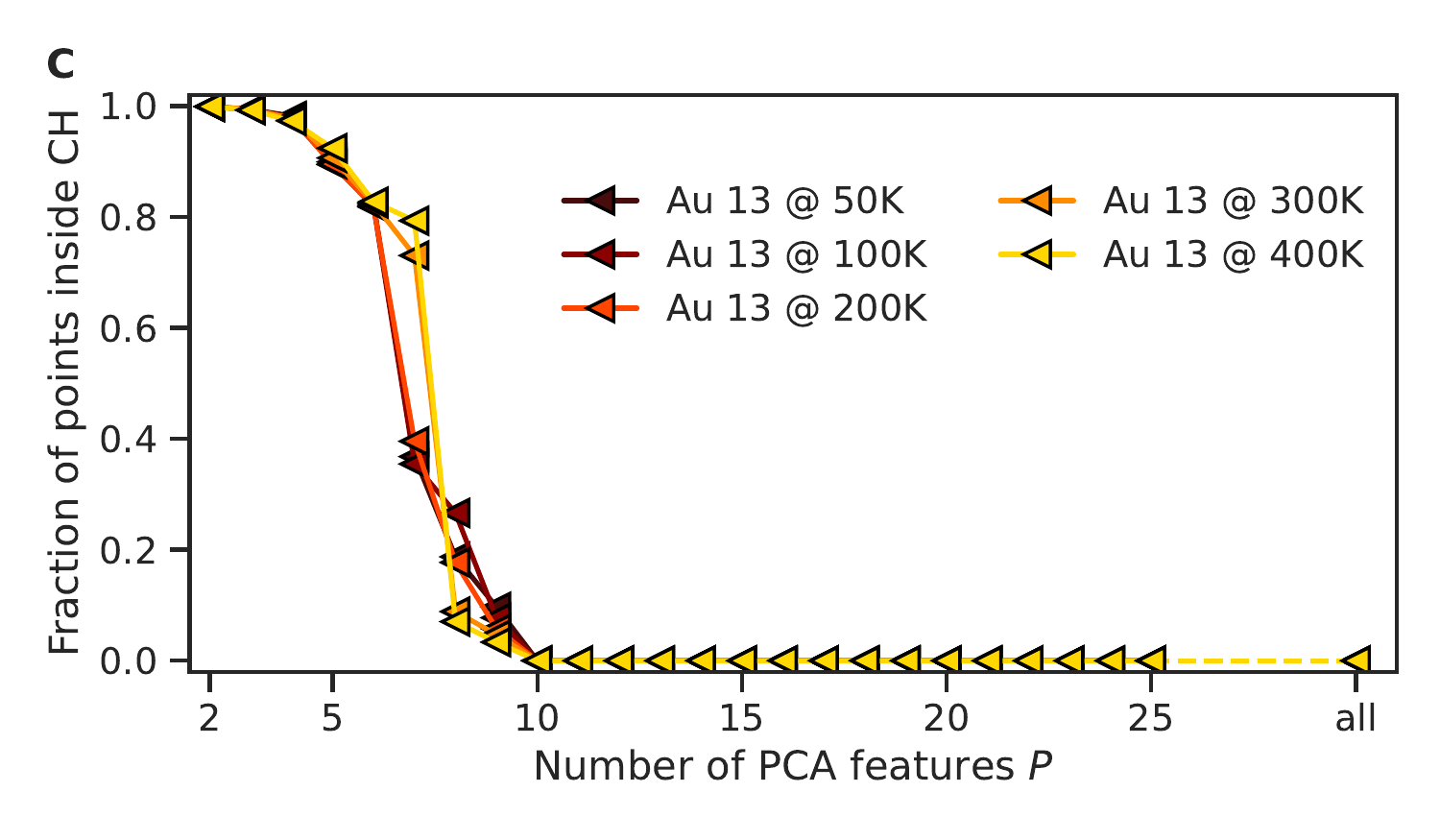}
    \caption{Fraction of atomic environments in the test set which are contained in the CH enclosing the points in the training set, featurized according to their first $P$ PCA components.
    Panel A refers to results found for the water-ice database, using SOAP or ACSF representations.
    Panel B corresponds to the case-study of the six datasets described in \citet{zuo2020performance}, using an ACE representation.
    Panel C reports results found for structures extracted from an MD trajectory of an Au$_{13}$ planar nanocluster at different temperatures, using an ACE representation, where the test and training set are constructed via a leave-one-out scheme. 
    }
    \label{fig:fig1}
\end{figure}

Figure \ref{fig:fig1} reports the number of test atomic environments which fall within the CH induced by the training set as a function of the dimensionality $P$ of the PCA representations, for the three databases described.
For reference we show in Figure S1 the cumulative variance explained by 2 to 25 principal components.
The test points are completely contained in the CH determined by the training set, when considering a projection in the space of the first two PCA components.
Increasing the dimensionality of the embedding, the number of test points enclosed within the CH diminishes rapidly.
Regardless of the database design, chemical nature of the system, and choice of representation (see also Figure~S2 for additional benchmarks), embedding on a low yet sizeable ($P\sim$10-20), number of PCA components results in an almost complete separation between each of the test points and the CH associated to the training points.

Following Figure \ref{fig:fig1}, we note that low-dimensional projections of atom-density features fail in faithfully preserving the information about whether an atomic environment is contained within the CH determined by a set of other ones.
At a more fundamental level, we highlight that ML potentials based on high-dimensional representations are very likely to carry out predictions at data points not contained in the convex-hull enclosing the training set.
This is true not only when testing the transferability of the ML potential from one phase to another (i.e., the ice-water database), but also in apparently trivial MD trajectory where no structural rearrangements take place (i.e., the \citet{zuo2020performance} and the Au$_{13}$ database).
Notably, the non-negligible portion of atomic environments found within the training points convex hull in the \citet{zuo2020performance} dataset (Cu, Ni, Ge and Si curves in Figure \ref{fig:fig1}B) actually consists of local atomic environments identical to those also present in the training set (See Supplemental Material, section D). 

\begin{figure}[t!]
    \centering
    \includegraphics[width=8.cm]{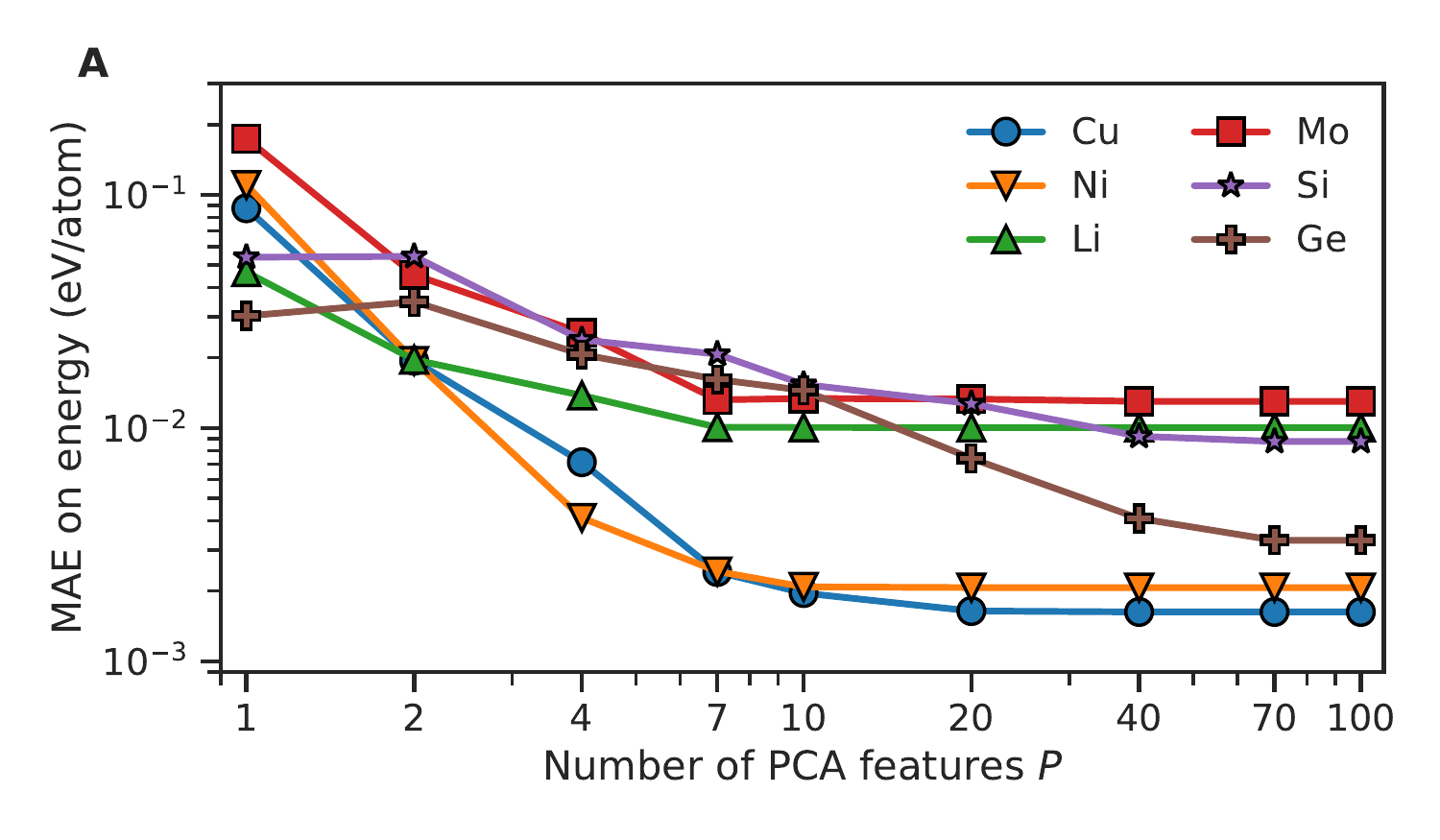}
    \includegraphics[width=8.cm]{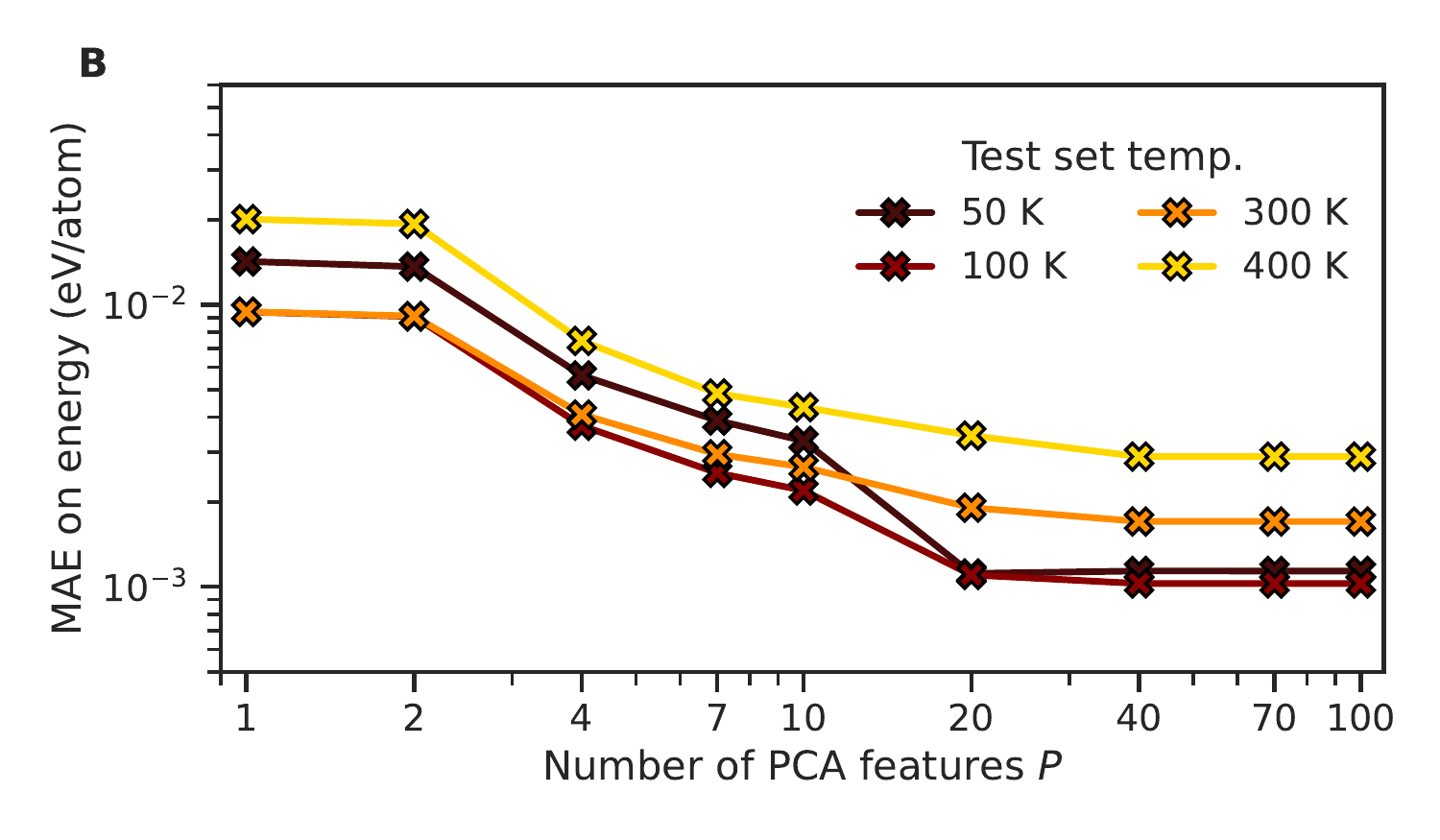}
    \caption{Mean Absolute Error on energy prediction when accounting for the contribution of the first $P$ PCA of the high-dimensional feature spaces deriving from an ACE representation for a ridge-regression potential trained on the \citet{zuo2020performance} dataset (panel A) and on the Au$_{13}$ database at 200~K (panel B).}
    \label{fig:fig2}
\end{figure}

After determining that predictions are likely to happen outside the training set convex hull when using a high-dimensional input representation, we test whether a ML potential's prediction effectively operates in such a high-dimensional space.
To investigate if an accurate ML potential projects data into a low-dimensional space where test data are contained by the ones in the training set, we train a regularized linear potential to predict the energy of structures for the example cases of the \citet{zuo2020performance} and the Au$_{13}$ datasets (see Supplemental Material, sections G and H for further info).
We analyze the weights assigned to each feature by the regression algorithm following training, and map the relevance of each PCA feature towards diminishing the per-atom energy mean absolute error (MAE).
We report the MAE of the potential as a function of the number of PCA components accounted for in the regression in Figure \ref{fig:fig2}. 
We refer the interested reader to Figure S1 for a report on the amount of variance explained, and to Figure \ref{fig:fig1} B-C for the number of test points which are enclosed by the training points CH, as a function of the first $P$ principal components.
We confirm that ridge regression potentials  do need to operate in a feature space where the majority of the test set atomic environments lie in a region of extrapolation to reach their best accuracy.

In light of the accuracy reported in the literature for predictions on train-set as well as on test-set configurations \cite{Monserrat2020, Cheng2019, zuo2020performance, Zeni2021descriptors} and our analysis, we demonstrate the need to revisit previous interpretations \cite{Musil2021, zuo2020performance, Monserrat2020, Bartok2018, Zeni2021,  Shao2021} relating ML potentials' accuracy and the geometrical relationship between test and training atomic environments, especially when these are carried out on low-dimensionality projections.
Indeed, ML potentials exploiting a high-dimensional local density representation can generalize their predictions to atomic environments which lie in an extrapolation regime, as per defined according to a convex-hull based criterion.
By the same token, and from the opposite perspective, we conclude that accurate predictions do not imply that test points are contained within the high-dimensional CH enclosing the set of available training points.
For the above reasons, a CH-based definition of interpolation and extrapolation is too weak and uninformative for the case of high-dimensional spaces.
This finding further motivates the search for better suited definitions of interpolation and extrapolation, which align with generalization performances.

To rationalize the effectiveness of ML regressors in high-dimensions we hypothesize that the test points lie in regions of the representation space that are sufficiently sampled by the training set distribution.
To test this hypothesis, we estimate the probability density generated by the training set points, which we call {\it sampling density},  on the locations of representation space where test points lie.
We do so by using an adaptive k-nearest-neighbor density estimation, which works as follows.
The test point $\mathbf{x^*}$ is virtually added to the training set and its $k^*$ nearest training points are found.
The density is then computed as
\begin{equation}
    \rho(\mathbf{x^*}) = \frac{k^* - 1}{M \, V^*},
\end{equation}
where $M$ is the size of the training set and $V^*$ is the volume occupied by the first $k^*$ training point neighbors. \cite{Carli2021}

The number $k^*$ of training neighbor considered is chosen adaptively for each test point as to maximize the accuracy of the estimate. \cite{Rodriguez2018}
The volume $V^*$ is computed as the volume of a hypersphere in dimension $d$, where $d$ is the intrinsic dimension of the training data manifold computed via the TwoNN estimator. \cite{Facco2017}
This way, we obtain a measure that indicates how much test set points are well-sampled by the training set density.
\
We refer the interested reader to Section I of the Supplemental Material for further detail on the density estimation algorithm.

\begin{figure}[t!]
    \centering
    \includegraphics[width=8.5cm]{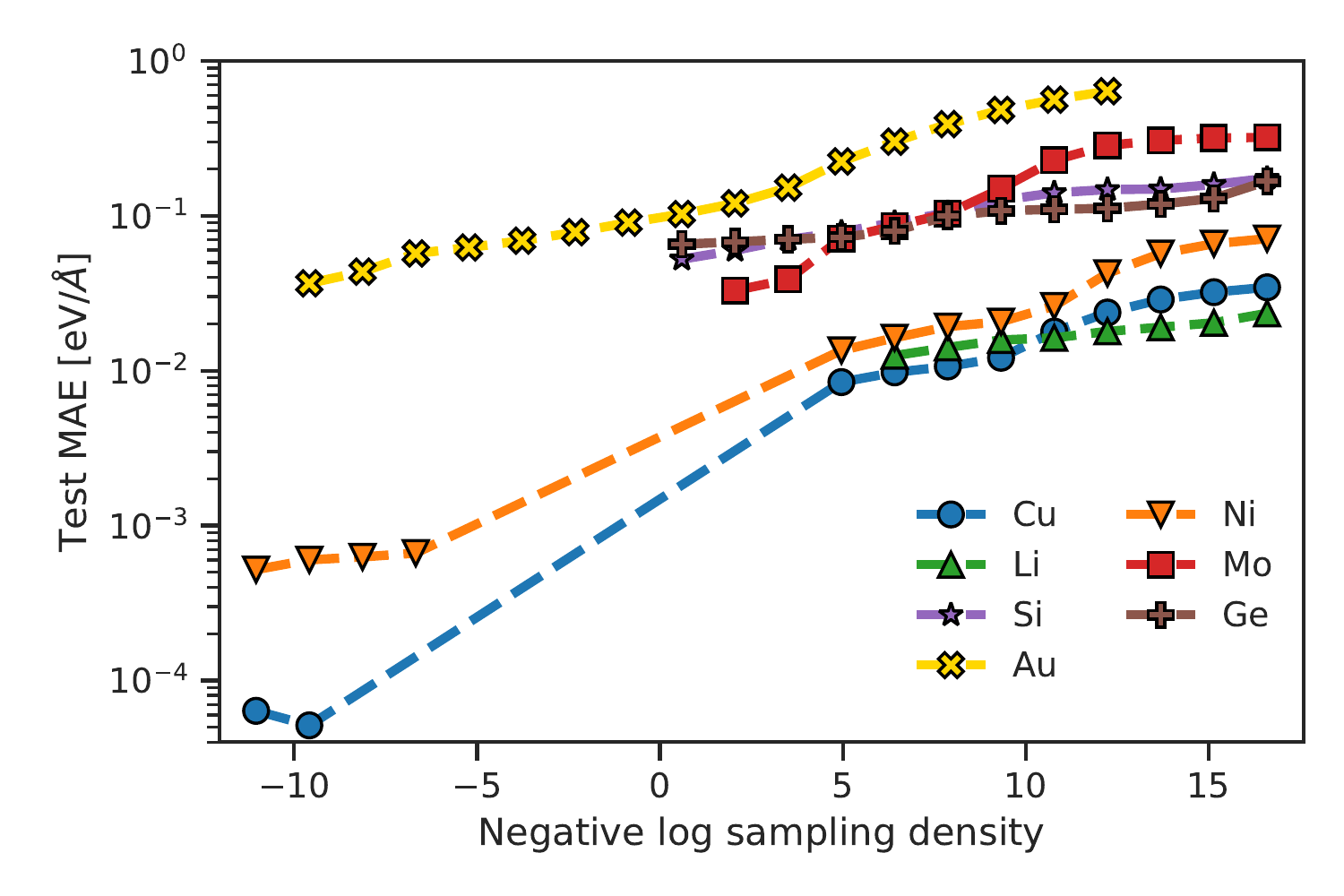}
    \caption{Correlation between the test MAE on forces incurred by ridge regression potentials, and the negative log probability density estimate for the training points in representation space, computed on test points (i.e., sampling density).
    }    
    \label{fig:fig3}
\end{figure}

We assess the relationship existing between such measure and the error incurred by a ridge-regression potential based on ACE representations trained on the data that generates the density distribution when tested on points in the test set.
We do so for the \citet{zuo2020performance} and for the Au$_{13}$ database.
In the latter case, each potential is trained on data from a MD trajectory at a single temperature, and tested on data coming from the other four MD trajectories.
In Figure \ref{fig:fig3}, we report the test MAE on forces, averaged and binned over the negative log sampling density of local atomic environment representations in the test sets.
We use 20 averaging windows equispaced between the lowest 5\% and the highest 5\% log density encountered across all test sets; and display only bins containing at least 1\% of the data.

Figure \ref{fig:fig3} highlights how the log sampling density correlates with the binned test MAE on forces for points outside the training set, for all databases considered (see also Figures S7-S9).
We find that the metric we introduce offers an estimate of the degree by which an out-of-sample atomic environment lies within a well-sampled region of the representation space and, more importantly, correlates with the error incurred by trained regression potentials.
We nevertheless find that the proposed metric is dependent on the choice of the representation (Figure S8-S9) and that the precise relationship with the MAE is system- and model-dependent.
\
Moreover, we observe a good correlation between the proposed metric and a model-dependent error estimator, i.e., the prediction uncertainty drawn from a committee of models trained by sub-sampling a larger training set \cite{Imbalzano2021} (See Section K, Figure S11 for further details).
These results confirm the hypothesis that the training set sampling density provides a statistical/geometrical criterion to chart extrapolation robustness of high-dimensional machine learning models across the representation space.

To conclude, we follow the definition spelled out by \citet{Balestriero2021}, where extrapolation occurs if the test structure lies outside the high-dimensional CH which encloses the set of training structure, and find
that the large majority of the test predictions take place in an extrapolation regime when employing high-dimensional local atomic density representations (ACSF, SOAP, ACE).
We thus demonstrate the need to revisit previous interpretations relating ML potentials' accuracy and the geometrical relationships between test and training atomic environments.

In second instance, to understand why linear ML models exploiting atom-density representations are predictive for points outside their convex-hull, we relate their accuracy to the probability density induced by training points in the representation space.
This geometric measure of well-sampledness in the representation space is found to correlate with the error incurred by the ML model, and overcomes the limitations in the use of a convex-hull construction to geometrically define interpolation and extrapolation regimes in high dimensional spaces.
The criterion suggested, in turn, enables to verify whether a training set is well-suited to enable accurate predictions at a target point in representation space.

We envision that the density-sampling analysis will promote the rational development of novel database generation routines, adaptive sampling protocols, data point selection algorithms.
This area of research \cite{Vandermause2020, Karabin2020, Bernstein2019, Podryabinkin2017} is indeed critical towards a data-efficient route in the automatic construction of machine learning potential potentials.
Future endeavours will be also directed to probe the mechanisms that rule the interpolation/extrapolation and the induced training densities inherent to ML potentials exploiting learnable representations \cite{Rupp2012, Schutt2018, Chmiela2019, Xie2018, Gilmer2017, Batzner2022}.
Finally, we highlight the generality of the training dataset sampling density analysis, which could be applied in other domains where high-dimensional ML models are widespread and successful, e.g., image recognition, diagnostics, therapeutic development and healthcare delivery.

\section*{Acknowledgements}
C.Z. gratefully acknowledges support by the European Union’s Horizon 2020 research and innovation program (Grant No. 824143, MaX `MAterials design at the eXascale' Centre of Excellence).
K.R. has received funding from the European Research Council (ERC) under the European Union’s Horizon 2020 research and innovation programme (Marie Curie Individual Fellowship Grant agreement No. 890414).
\\
The authors thank
Matteo Carli,
Stefano De Gironcoli,
Piero Gasparotto, 
Federico Grasselli,
Andrea Grisafi, 
Giulio Imbalzano, and
Nataliya Lopanitsyna 
for useful discussions. 
\\
In memory of Alessandro De Vita.

\section{Supplemental Material}
See Supplemental Material for additional details regarding the local atomic environment representations employed, the generation of the Au$_{13}$ dataset, additional CH benchmarks, the effect of identical local atomic environments on CH computation, ridge regression for energies and forces, errors as a function of PCA components in ridge regression for the $Au_{13}$ database, the adaptive k-NN estimation algorithm, the correlation between the negative log density and the MAE incurred by ridge regression force fields, and the comparison of adaptive k-nn with other error estimation methods.

\section*{Data availability}
Scripts and inputs required to generate all the plots and results discussed in this article are available on the Materials Cloud (DOI 10.24435/materialscloud:8w-a7)
\section*{Conflict of interest}
 
A.A. is currently employed at Hoffman La Roche and has contributed to this work outside of his Roche Post Doctoral Fellowship project. 
 
\section*{Bibliography}

\bibliography{main_arxiv}

\clearpage
\newpage

\section*{Supplemental Material}

\renewcommand{\thetable}{S\arabic{table}}
\renewcommand{\thefigure}{S\arabic{figure}}
\setcounter{figure}{0}


\subsection{Representation for atomistic systems}
In the context of ML potentials, configuration of atoms can be informatively expressed by means of atomic environment representations.
The atomic environment representation for atom $i$ encodes information about the position and atomic species of all atoms $j$ contained in a ball of radius  $r_{c}$ centered on $i$.
In this investigation, we considered three frameworks to represents atomic environments that found a wide-spread and successful use in the literature.
A brief resume of their salient characteristic is discussed in detail in the next corresponding subsections.

\subsubsection{Atom-centered Symmetry Functions representation}

Atom-centered Symmetry Functions representation (ACSF) encode information about the arrangement of atoms within an atomic environment through a discrete set of features which map 2- and 3-body correlations.
Behler-Parrinello \cite{Behler:2007fe} Radial symmetry functions $G_2^s$ are calculated as sums of Gaussians of width $1/\eta$ and centered at $r_s$:
\begin{equation}
    G_2^s = \sum_{j | r_{ij} \leq r_c} e^{-\eta (r_{ij}-r_s)^2} f_c(r_{ij}).
\end{equation}
Angular symmetry functions are built as sums of cosine functions of the angles $\theta_{ijk} = acos ( \frac{ {\bf r_{ij} \dot r_{ik}} }{r_{ij} \dot r_{ik} } ) $
:
\begin{equation}
\begin{aligned}
        G_3 = 2^{1-\xi} \sum_{j, k | r_{ij}, r_{ik} \leq r_c}  (1 + \lambda cos(\theta_{ijk}) )^\xi \\
    e^{-\eta (r_{ij}^2 + r_{ik}^2 + r_{jk}^2)} 
    f_c(r_{ij}) f_c(r_{ik}) f_c(r_{jk}),
\end{aligned}
\end{equation}
where $\lambda$, whose value ranges between 1 and -1, tunes the maxima of the cosine function, and $\xi$ tunes the angular resolution.
We compute ACSF features using the n2p2 package. \cite{Singraber2019}

\subsubsection{Atom density representations}
Atom-density representations describe the atomic density $\rho_i(\mathbf{r})$ as a sum of functions $u(\mathbf{r}_{ji} - \mathbf{r})$ centered on each atom $j$ within a cut-off $r_{c}$ around the atom $i$:
\begin{equation}
\rho_i(\mathbf{r}) = \sum_{j | r_{ij} \leq r_c}  u (\mathbf{r}_{ji} - \mathbf{r}),
\label{eq:local_atomic_environment}
\end{equation}
where $\mathbf{r}_{ji}$ indicates the vector $(\mathbf{r}_j - \mathbf{r}_i)$.

Under the Smooth Overlap of Atomic Positions (SOAP) representation framework,\cite{Bartok2010} $u(\mathbf{r}_{ji} - \mathbf{r})$ takes the form of a Gaussian of width $\sigma$:
\begin{equation}
        \rho_i(\mathbf{r})  =\ \sum_{j | r_{ij} \leq r_c} e^{\left(-\frac{\|\mathbf{r}- \mathbf{r}_i\|^2}{\sigma^2}\right)}.
\end{equation}

Under the Atomic Cluster Expansion (ACE) representation framework,\cite{Drautz2019}
Dirac delta functions are employed in place of Gaussians:
\begin{equation}
    \rho_i(\mathbf{r}) = \sum_{j | r_{ij} \leq r_c} \delta(\mathbf{r}_{ji} - \mathbf{r}),
\end{equation}

The atomic environment representations are then approximated via a truncated expansion in spherical harmonics and radial basis functions.
The atomic density $\rho_i(\mathbf{r})$ therefore takes the form:
\begin{equation}
\rho_i(\mathbf{r}) \sim \sum_{j \in \rho_i} \sum_{n=0}^{n_{MAX}} \sum_{l=0}^{l_{MAX}} \sum_{m=-l}^{l} c_{nlm}^j f_{n}(r_{ji}) Y_{lm}(\hat{\mathbf{r}}_{ji}),
\label{eq:spherical_harmonics_expansion}
\end{equation}
where $\hat{\mathbf{r}}_{ji}$ is the unit vector of $\mathbf{r}_{ji}$, $f_{n}$ are elements of a set of radial basis functions, $Y_{lm}$ are elements of a set of spherical harmonics, $n_{MAX}$ indicates the truncation limit for the radial basis set, and $l_{MAX}$ the truncation limit for the angular basis set.
Finally, rotationally invariant representations are built as products of $N$ coefficients $c_{nlm}^j$ that correspond to a reducible representation of the identity of the rotation group.
The resulting representations are of order $(N+1)$, i.e. can encode the interaction of up to $N+1$ atoms at once. \cite{Drautz2019, Glielmo2018}
We compute SOAP features using the DScribe Python package \cite{dscribe}, while ACE features are computed using a custom-made Python wrapper for the ACE.jl package presented in \citet{Dusson2022}.

\newpage

\subsection{Representation parameters}

Table \ref{tab:tab0} gathers the SOAP representation parameters employed for the \citet{Monserrat2020} dataset, and mirror the ones reported in the original work.
\begin{table}[h!]
    \centering
    \begin{tabular}{c||c|c|c|c}
        system & ~~~ r$_{c}$ ~~~ & ~~ n$_{max}$ ~~ & ~~ l$_{max}$  ~~ & ~~ $\sigma$ \\
        \hline
        \hline
        H$_{2}$O & 6 & 6 & 6 & 0.5 \\
    \end{tabular}
    \caption{Resume of the SOAP representation parameters adopted to expand atom-densities around atom centred environments in the \citet{Monserrat2020} database.}
    \label{tab:tab0}
\end{table}

Tables \ref{tab:tab00} and \ref{tab:tab000}, respectively, gather the 2- and 3-body ACSF parameters employed for the \citet{Monserrat2020} dataset, and mirror the ones reported in the original work.
\begin{table}[h!]
    \centering
    \begin{tabular}{c||c|c|c|c}
central atom & neighbour atom & ~~ $\eta$ ~~ & ~~ $r_s$ ~~ & ~~ $r_c$ ~~ \\
\hline
\hline
 H & H & 0.001 &  0.0 &  12.00\\
 H & H &  0.01 &   0.0 &  12.00  \\
 H & H &  0.03 &   0.0 &  12.00  \\
 H & H &  0.06 &   0.0 &  12.00  \\
 H & H &  0.15 &   1.9 &  12.00  \\
 H & H  & 0.30 &   1.9 &  12.00  \\
 H & H  & 0.60 &   1.9 &  12.00 \\
 H & H  & 1.50 &   1.9 &  12.00  \\
 H & O  & 0.001 &  0.0 &  12.00    \\
 H & O &  0.01 &   0.0 &  12.00   \\
 H & O &  0.03 &   0.0 &  12.00   \\
 H & O &  0.06 &   0.0 &  12.00    \\
 H & O &  0.15 &   0.9 &  12.00 \\
 H & O &  0.30 &   0.9 &  12.00   \\
 H & O &  0.60 &   0.9 &  12.00    \\
 H & O &  1.50 &   0.9 &  12.00   \\
 O & H &  0.001 &  0.0 &  12.00   \\
 O & H &  0.01 &   0.0 &  12.00    \\
 O & H &  0.03 &   0.0 &  12.00   \\
 O & H &  0.06 &   0.0 &  12.00   \\
 O & H &  0.15 &   0.9 &  12.00   \\
 O & H &  0.30 &   0.9 &  12.00   \\
 O & H &  0.60 &   0.9 &  12.00  \\
 O & H &  1.50 &   0.9 &  12.00   \\
 O & O &  0.001 &  0.0 &  12.00  \\
 O & O &  0.01 &   0.0 &  12.00 \\
 O & O &  0.03 &   0.0 &  12.00   \\
 O & O &  0.06 &   0.0 &  12.00\\
 O & O &  0.15 &   4.0 &  12.00 \\
 O & O &  0.30 &   4.0 &  12.00  \\
 O & O &  0.60 &   4.0 &  12.00  \\
 O & O &  1.50 &   4.0  & 12.00   \\    
    \end{tabular}
    \caption{Resume of the 2-body ACSF representation parameters adopted to expand atom-densities around atom centred environments in the \citet{Monserrat2020} database.}
    \label{tab:tab00}
\end{table}

\newpage

\begin{table}[h!]
    \centering
    \begin{tabular}{c||c|c|c|c|c|c}
central atom & neighbour 1 & neighbour 2 & $\eta$ & $\lambda$ & $\xi$ & $r_c$\\
\hline
\hline
H & O & H & 0.2   &  1.0 & 1.0 & 12.00   \\   
O & H & H & 0.07  & 1.0  & 1.0 & 12.00  \\    
H & O & H & 0.07  & 1.0  & 1.0 & 12.00  \\    
O & H & H & 0.07  & -1.0 & 1.0 & 12.00  \\    
H & O & H & 0.07  & -1.0 & 1.0 & 12.00  \\    
O & H & H & 0.03  & 1.0  & 1.0 & 12.00  \\    
H & O & H & 0.03  & 1.0  & 1.0 & 12.00  \\    
O & H & H & 0.03  & -1.0 & 1.0 & 12.00  \\    
H & O & H & 0.03  & -1.0 & 1.0 & 12.00  \\    
O & H & H & 0.01  & 1.0  & 4.0 & 12.00  \\    
H & O & H & 0.01  & 1.0  & 4.0 & 12.00  \\    
O & H & H & 0.01  & -1.0 & 4.0 & 12.00  \\    
H & O & H & 0.01  & -1.0 & 4.0 & 12.00  \\    
O & O & H & 0.03  & 1.0  & 1.0 &  12.00   \\   
O & O & H & 0.03  & -1.0 & 1.0 &  12.00   \\   
O & O & H & 0.001 &  1.0 & 4.0 & 12.00  \\    
O & O & H & 0.001 & -1.0 & 4.0 & 12.00   \\   
H & O & O & 0.03  & 1.0  & 1.0 &  12.00  \\    
H & O & O & 0.03  & -1.0 & 1.0 &  12.00  \\    
H & O & O & 0.001 &  1.0 & 4.0 & 12.00  \\    
H & O & O & 0.001 & -1.0 & 4.0 & 12.00  \\    
O & O & O & 0.03  & 1.0  & 1.0 &  12.00  \\   
O & O & O & 0.03  & -1.0 & 1.0 &  12.00  \\    
O & O & O & 0.001 &  1.0 & 4.0 & 12.00  \\    
O & O & O & 0.001 & -1.0 & 4.0 & 12.00  \\    
\\    
    \end{tabular} 
    \caption{Resume of the 3-body ACSF representation parameters adopted to expand atom-densities around atom centred environments in the \citet{Monserrat2020} database.}
    \label{tab:tab000}
\end{table}

A resume of the SOAP and ACE parameters for the case of the \citet{zuo2020performance} dataset are reported in Table \ref{tab:tab2} and \ref{tab:tab3}.
\begin{table}[h!]
    \centering
    \begin{tabular}{c||c|c|c|c}
        element & ~~~ r$_{c}$ ~~~ & ~~ n$_{max}$ ~~ & ~~ l$_{max}$  ~~ & ~~ $\sigma$ \\
        \hline
        \hline
        Cu & 3.9 & 8 & 8 & 0.5 \\
        Ni & 3.9 & 8 & 8 & 0.5 \\
        Mo & 5.2 & 8 & 8 & 0.5 \\
        Li & 4.8 & 8 & 8 & 0.5 \\
        Si & 5.4 & 8 & 8 & 0.5 \\
        Ge & 5.4 & 8 & 8 & 0.5 \\
    \end{tabular}
    \caption{Resume of the SOAP representation parameters adopted to expand atom-densities around atom centered environments in the \citet{zuo2020performance} database.}
    \label{tab:tab2}
\end{table}

\begin{table}[h!]
    \centering
    \begin{tabular}{c||c|c|c}
        element & ~~~ r$_{c}$ ~~~ & ~~ n$_{max}$ + l$_{max}$  ~~ & ~~ $N$ \\
        \hline
        \hline
        Cu & 3.9 & 10 & 5 \\
        Ni & 3.9 & 10 & 5 \\
        Mo & 5.2 & 10 & 5 \\
        Li & 4.8 & 10 & 5 \\
        Si & 5.4 & 10 & 5 \\
        Ge & 5.4 & 10 & 5 \\
    \end{tabular}
    \caption{Resume of the ACE representation parameters adopted to expand atom-densities around atom centered environments in the \citet{zuo2020performance} database.}
    \label{tab:tab3}
\end{table}

\newpage

Table \ref{tab:au2} recaps the ACE representation parameters employed for the Au$_{13}$ dataset.
\begin{table}[h!]
    \centering
    \begin{tabular}{c||c|c|c}
        element & ~~~ r$_{c}$ ~~~ & ~~ n$_{max}$ + l$_{max}$  ~~ & ~~ $N$ \\
        \hline
        \hline
        Au & 5 & 10 & 5 \\
    \end{tabular}
    \caption{Resume of the ACE representation parameters adopted to expand atom-densities around atom centered environments for the Au configurations.}
    \label{tab:au2}
\end{table}

\subsection{Au$_{13}$ DATASET GENERATION DETAILS}
Data is gathered from MD simulations of an Au$_{13}$ nanocluster ran for 3~ns (configurations are stored every 3~ps, for a total of 1000 de-correlated structures) at temperatures of 50~K, 100~K, 200~K, 300~K, and 400~K.
The initial frame of each trajectory corresponds to the relaxed global minimum structure for Au$_{13}$.
To simulate the trajectories, we integrate the dynamics with a 3~fs timestep, enforce an NVT ensemble via a velocity rescaling thermostat, with a time-scale of 200~fs, model interatomic interactions within the Density functional tight binding framework \cite{B.Aradi2007}, and use i-PI \cite{Kapil2019} as the MD driver.
\\
\newpage

\subsection{PCA explained variance}

\begin{figure}[h!]
    \centering
    \includegraphics[width=8cm]{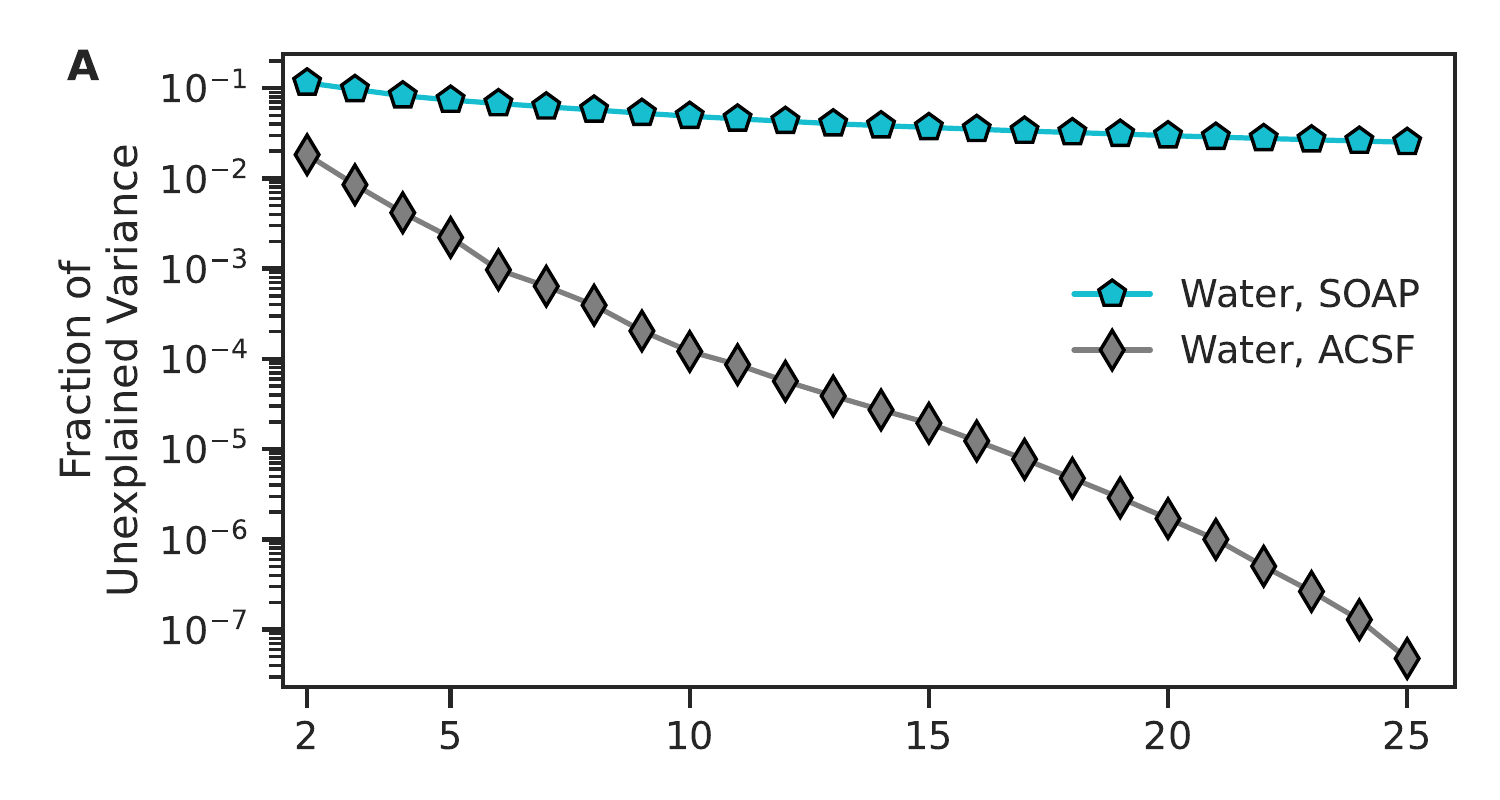}
    \includegraphics[width=8cm]{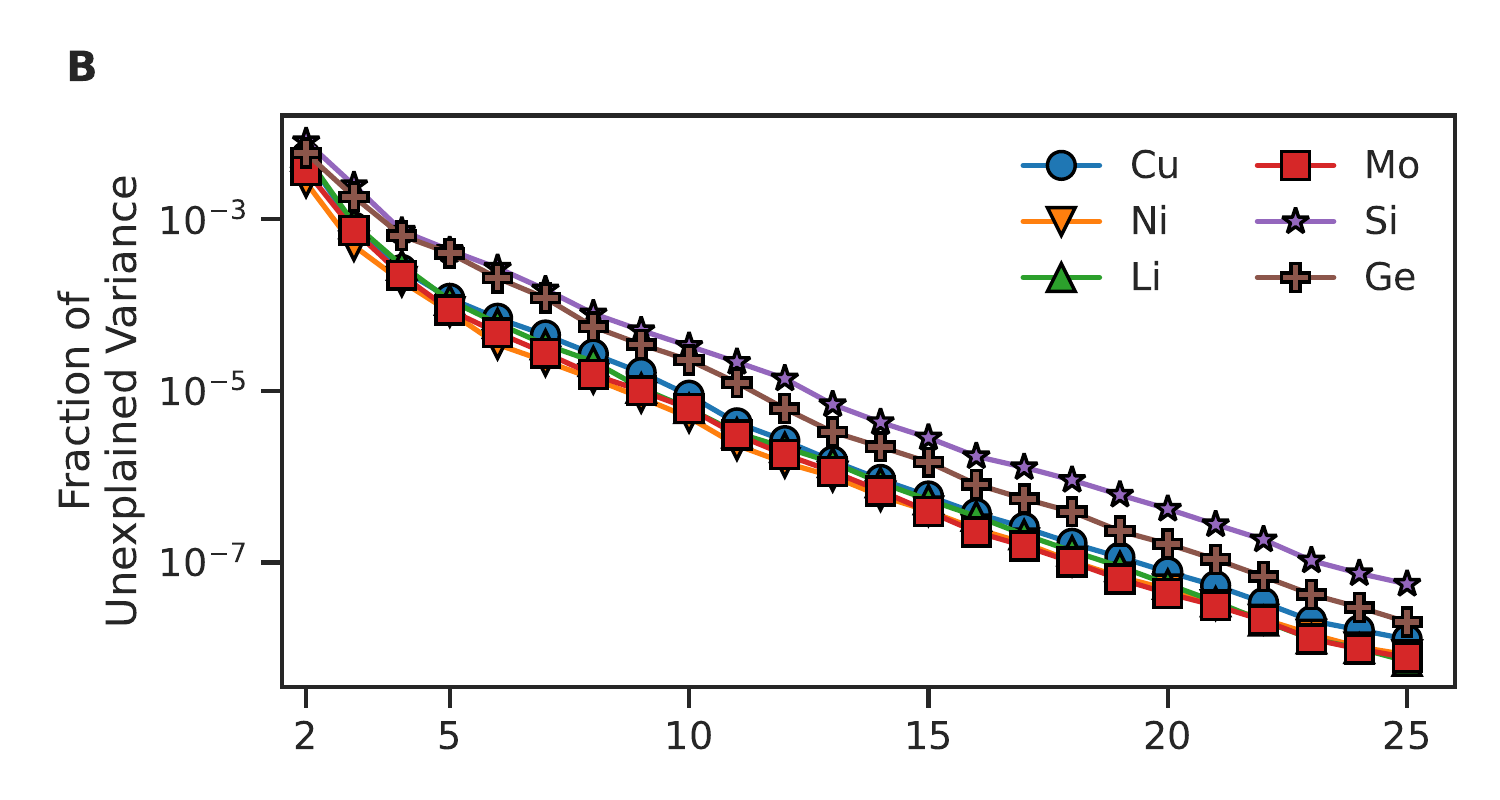}
    \includegraphics[width=8cm]{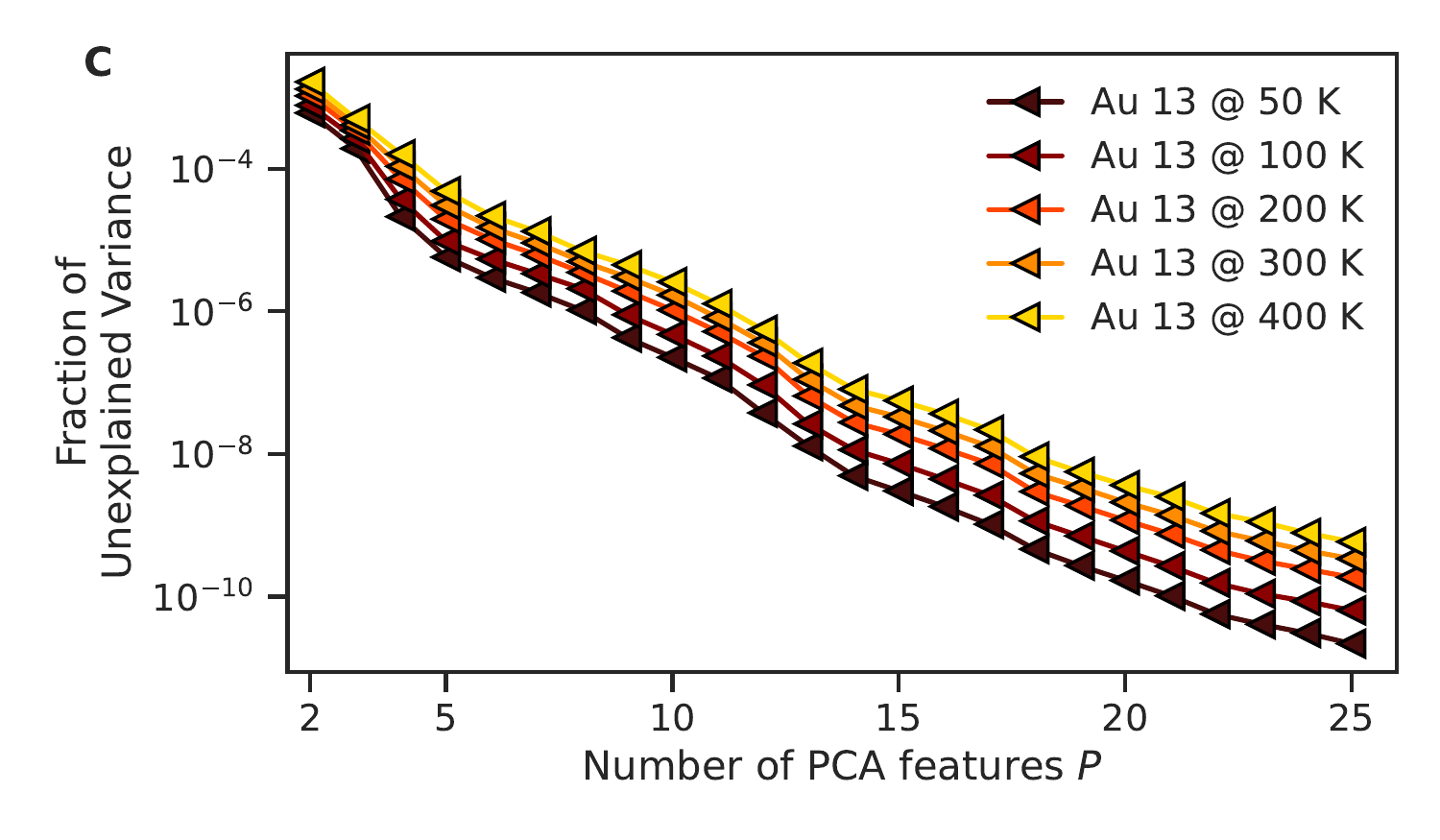}
    \caption{Fraction of unexplained variance in the data explained by the first $P$ PCA for feature spaces deriving from the system-dependent atom-density representations. 
    Data relative to both the SOAP (cyan pentagons) and ACSF (grey diamonds) features are displayed in Panel A.
    Panels B and C report ACE features with $n_{max} + l_{max} = 10$ and $N=5$.
    }
    \label{fig:pca-variance}
\end{figure}

\newpage
\newpage

\subsection{Additional Convex Hull Benchmarks}
We report about tests regarding the number of training points inside the CH defined by the test set for the case of Ni, Li, and Si databases when exploiting a SOAP and an ACE representation with a different number of radial and angular basis set components w.r.t. the ones employed in the main text. 
In particular, Figure S2 reports the same plots as in Figure 1B, but for ACE representations with  n$_{max}$ + l$_{max}$ = 6, 8, and for SOAP representations with  n$_{max}$, l$_{max}$ = 4, 6, and 8.
We find that the number of test points within the convex hull enclosing the training set is similar across the different SOAP and ACE set-ups.
This result holds both for the analysis in the full-dimensional space and in the low-dimensional PCA embedding.

\begin{figure}[h!]
    \centering
    \includegraphics[width=7cm]{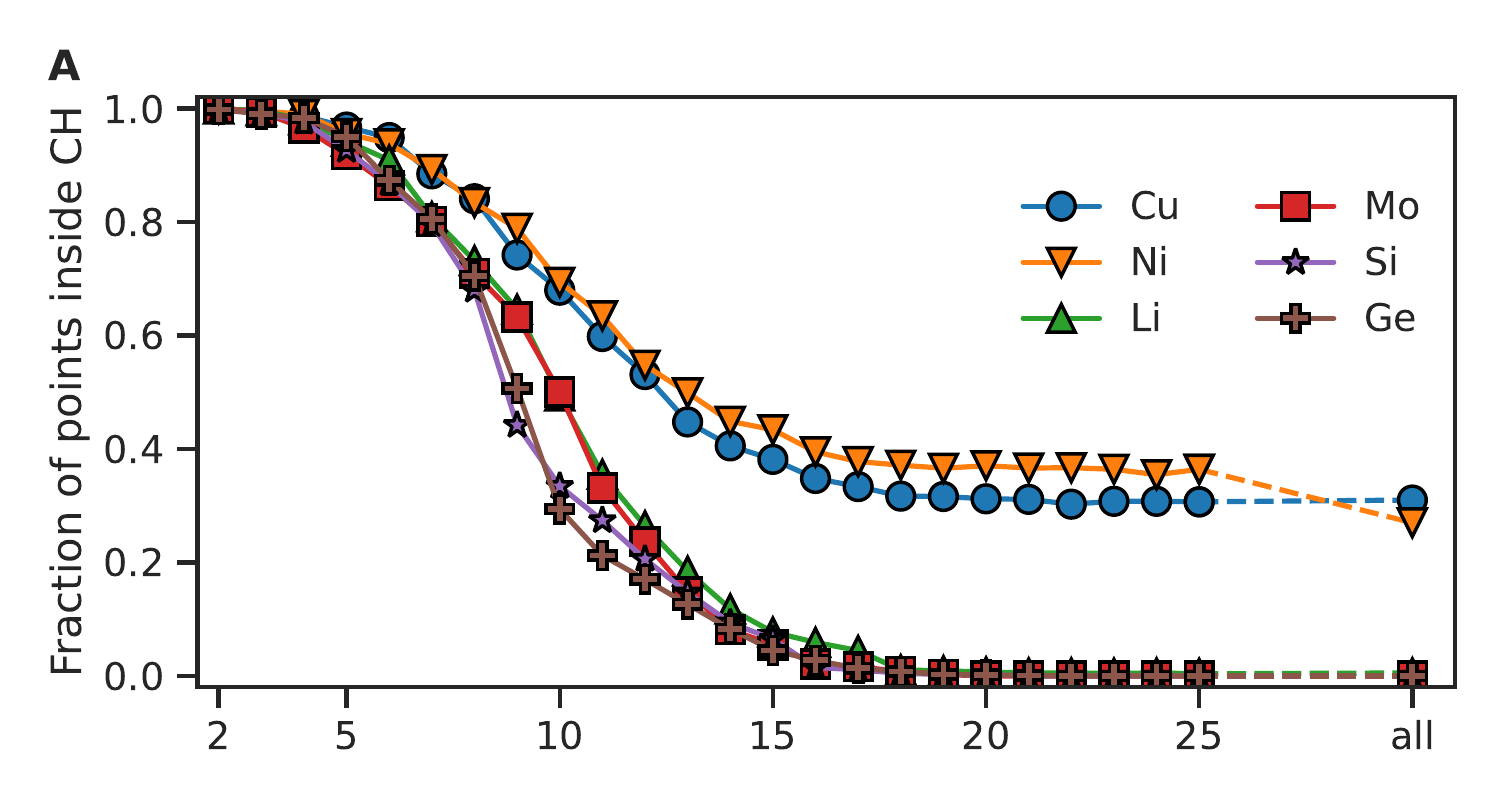}
    \includegraphics[width=7cm]{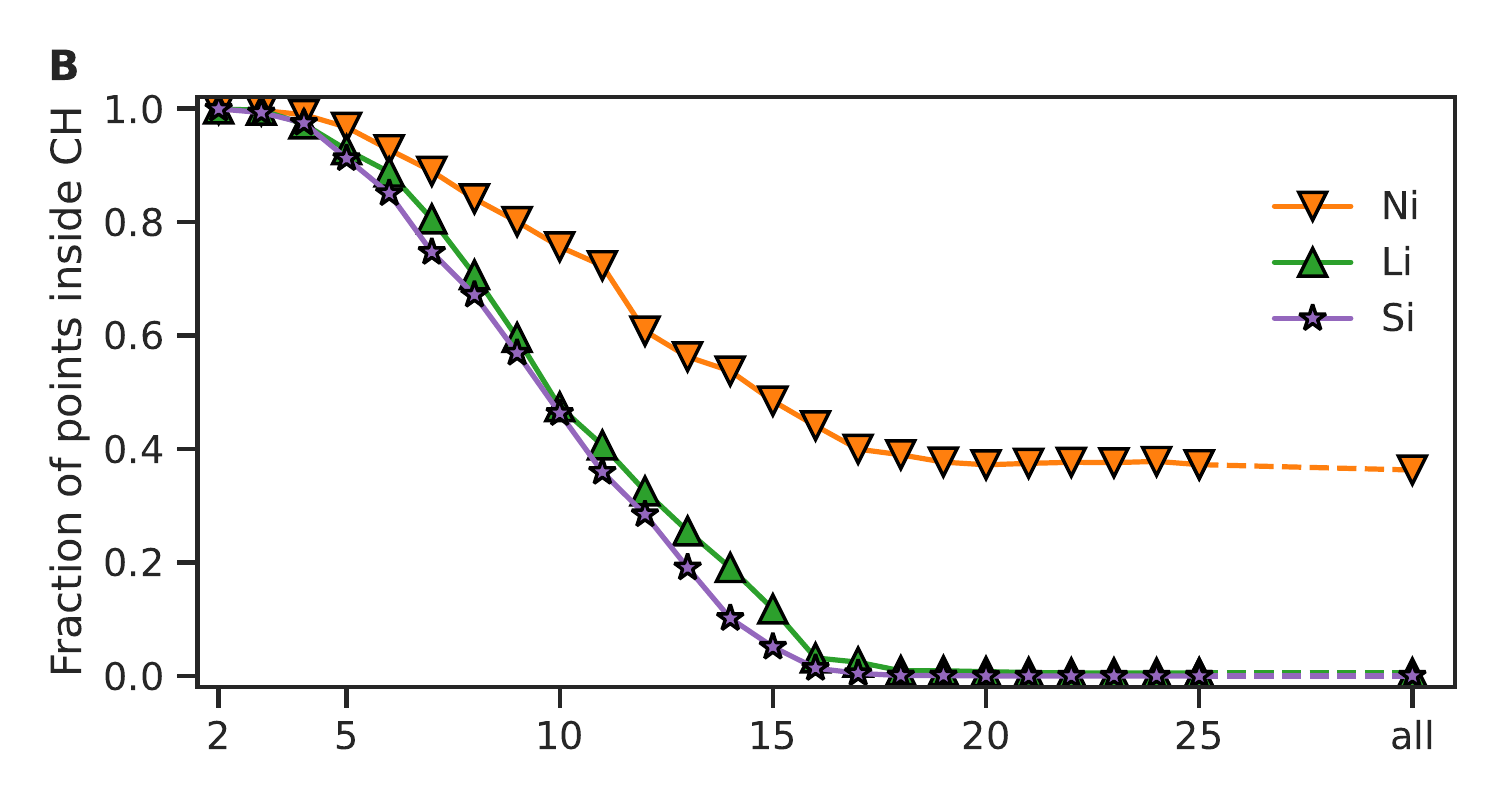}
    \includegraphics[width=7cm]{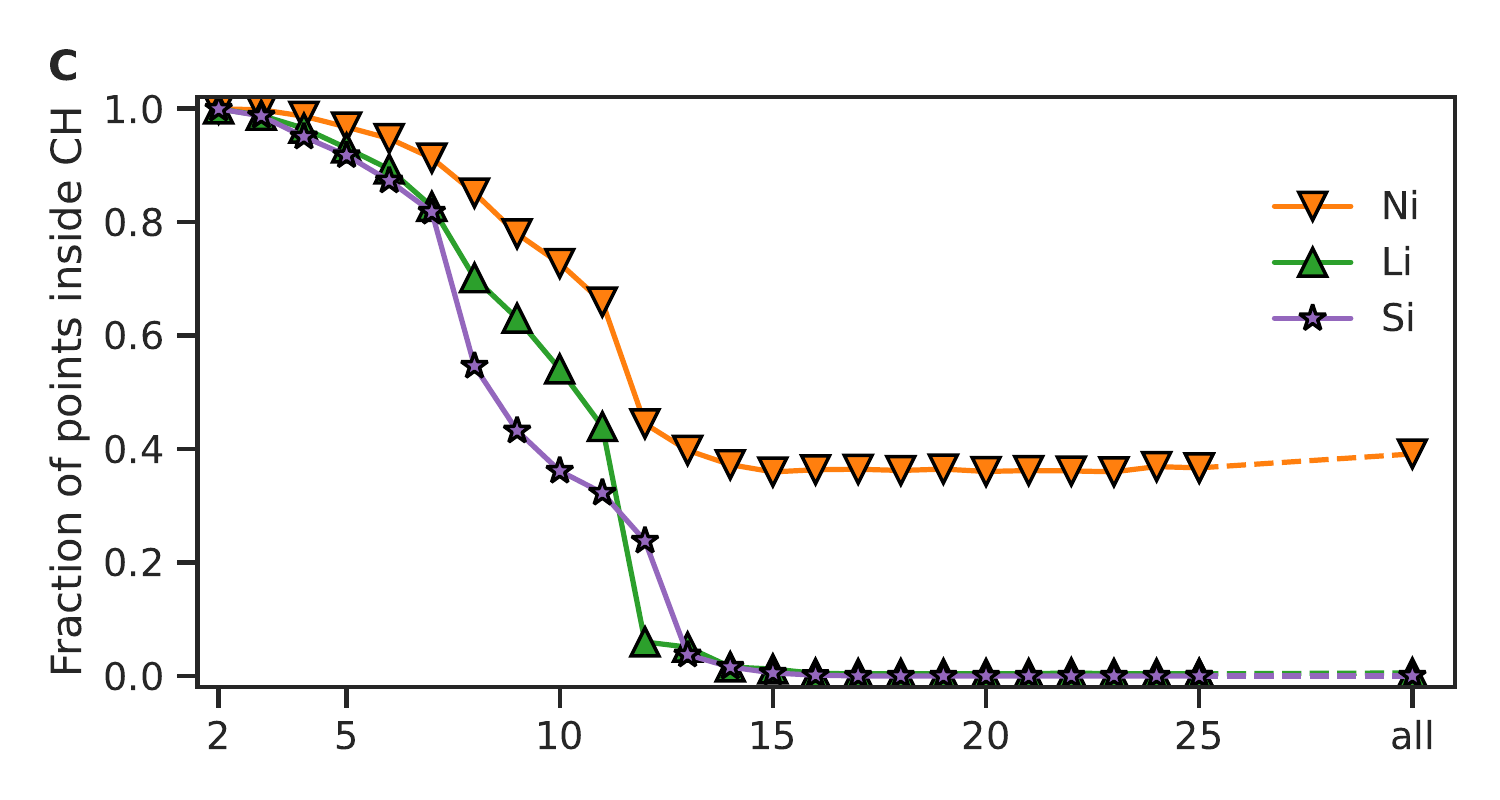}
    \includegraphics[width=7cm]{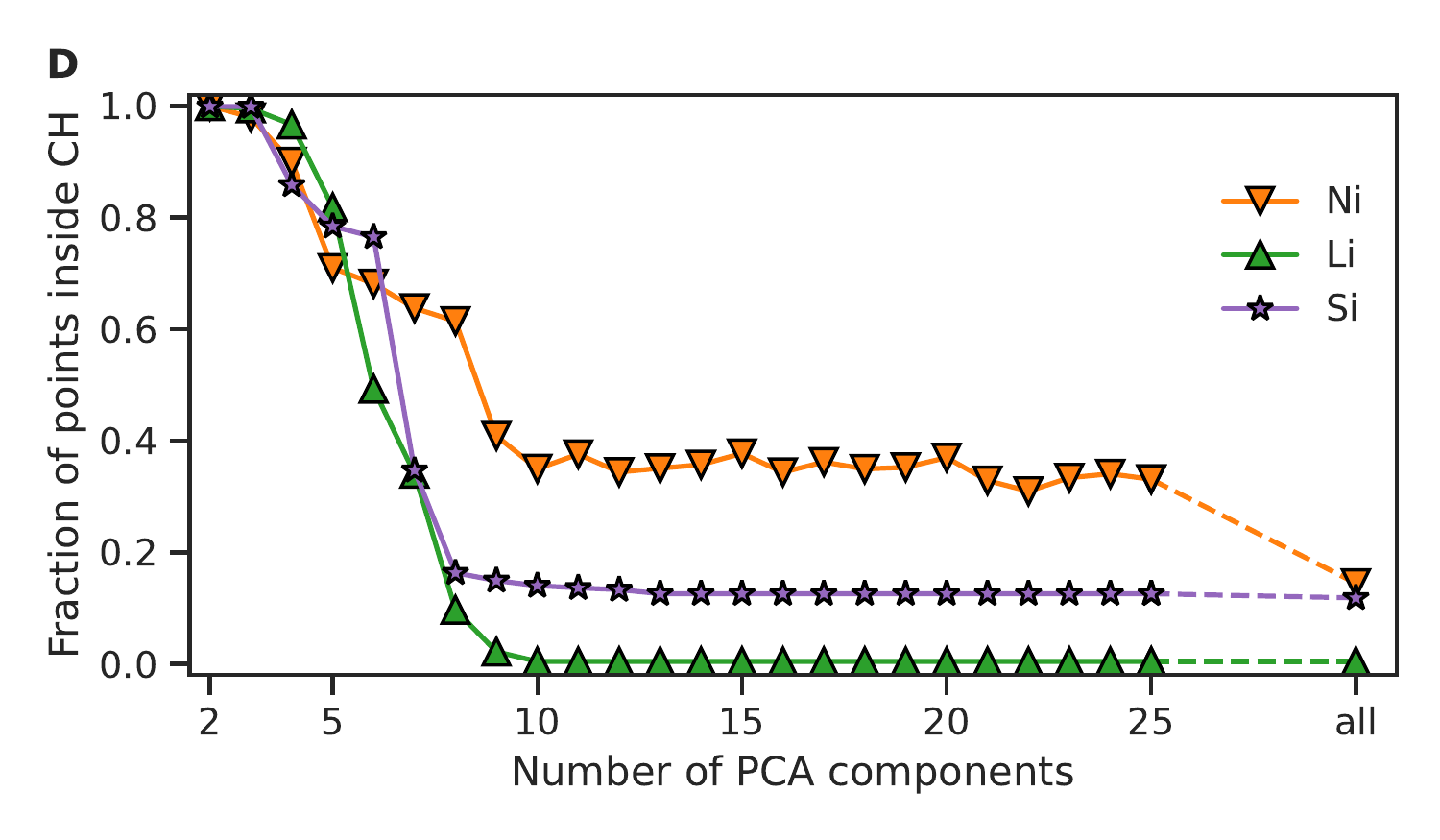}
    \includegraphics[width=7cm]{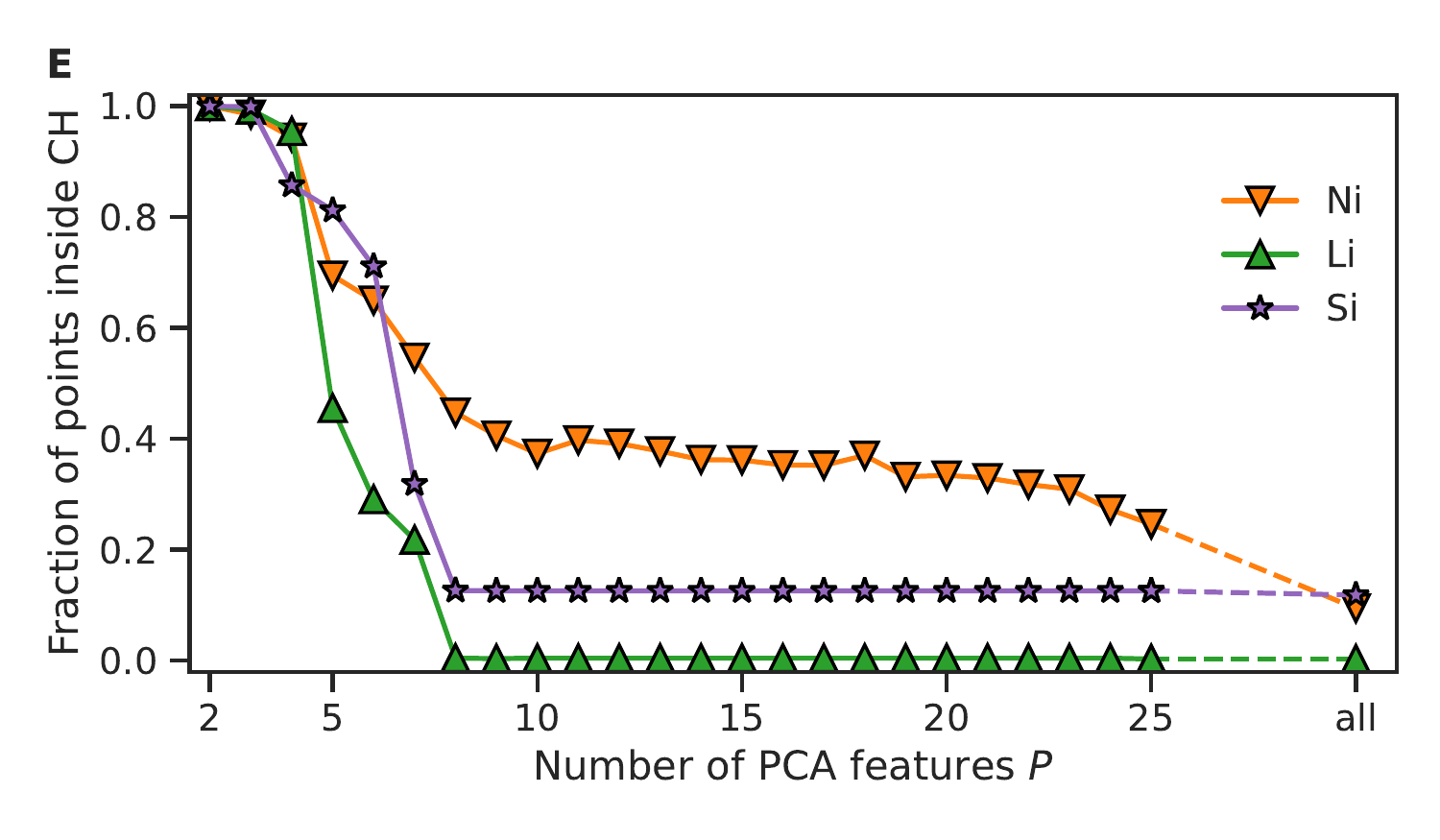}
    \caption{Fraction of atomic environments in the test set which are contained in the N-dimensional CH enclosing the points in the training set, featurized according to the first $P$ PCA components of high dimensional features.
    Panel A: SOAP representation using n$_{max}$ =8, l$_{max}$ =8.
    Panel B: SOAP representation using n$_{max}$ =6, l$_{max}$ =6.
    Panel C: SOAP representation using n$_{max}$ =4, l$_{max}$ =4.
    Panel D: ACE representation using  n$_{max}$ +l$_{max}$ =8.
    Panel E: ACE representation using n$_{max}$ +l$_{max}$ =6.}
    \label{fig:nltest}
\end{figure}

\newpage
\clearpage

\subsection{Effect of identical atomic environments on the convex hull computation}
Figure S3 shows the relative abundance of Euclidean distances between ACE representations in each \citet{zuo2020performance} test and training set.
For the case of the Cu, Ni, Si, and Ge datasets, respectively 6\%, 7\%, 18\% and 17\% of such distances are smaller than 10$^{-12}$, whereas this value goes to less than 0.1\% for the Li and Mo datasets.
We remark that identical atomic environments can be found in different configurations; this happens e.g., when comparing atoms in relaxed crystalline bulk with atoms in relaxed surfaces.

\begin{figure}[h!]
    \centering
    \includegraphics[width=8cm]{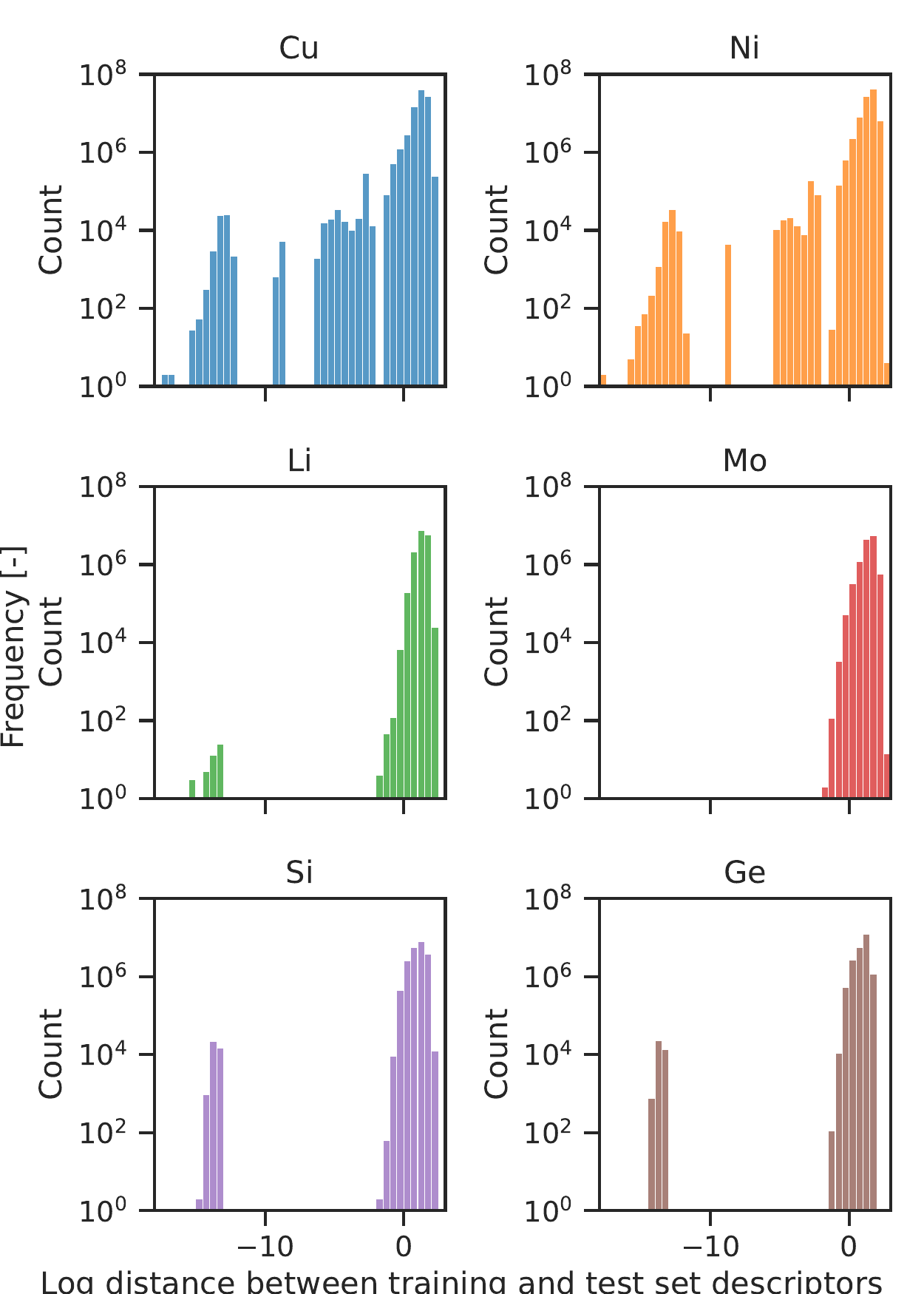}
    \caption{Histogram of distances between ACE representations of atomic environments in the training and test sets.
    The ACE representation employs $N$ = 5 and n$_{max}$ + l$_{max}$ = 10 as hyper parameters.}
    \label{fig:my_label}
\end{figure}

\newpage

Figure S4 highlights that test atomic environments for which exists an almost identical atomic environment in the training set are likely to be counted as within the CH defined by the training set.
When this anomalous atomic environments are not accounted for, the fraction of points inside the CH rapidly and monotonically goes to 0 as the number of PCA components increases.
This is even clearer in Figure S5, which shows the fraction of test ACE representations inside the CH defined by the training ACE representation for test representations with no distance smaller than 10$^{-12}$ from test representations.
We observe that, once we remove from the test set environments that are almost identical to ones in the training set, the fraction of points inside the CH goes consistently to zero for all six elemental datasets at around 15 PCA components.

\begin{figure}[h!]
    \centering
    \includegraphics[width=8cm]{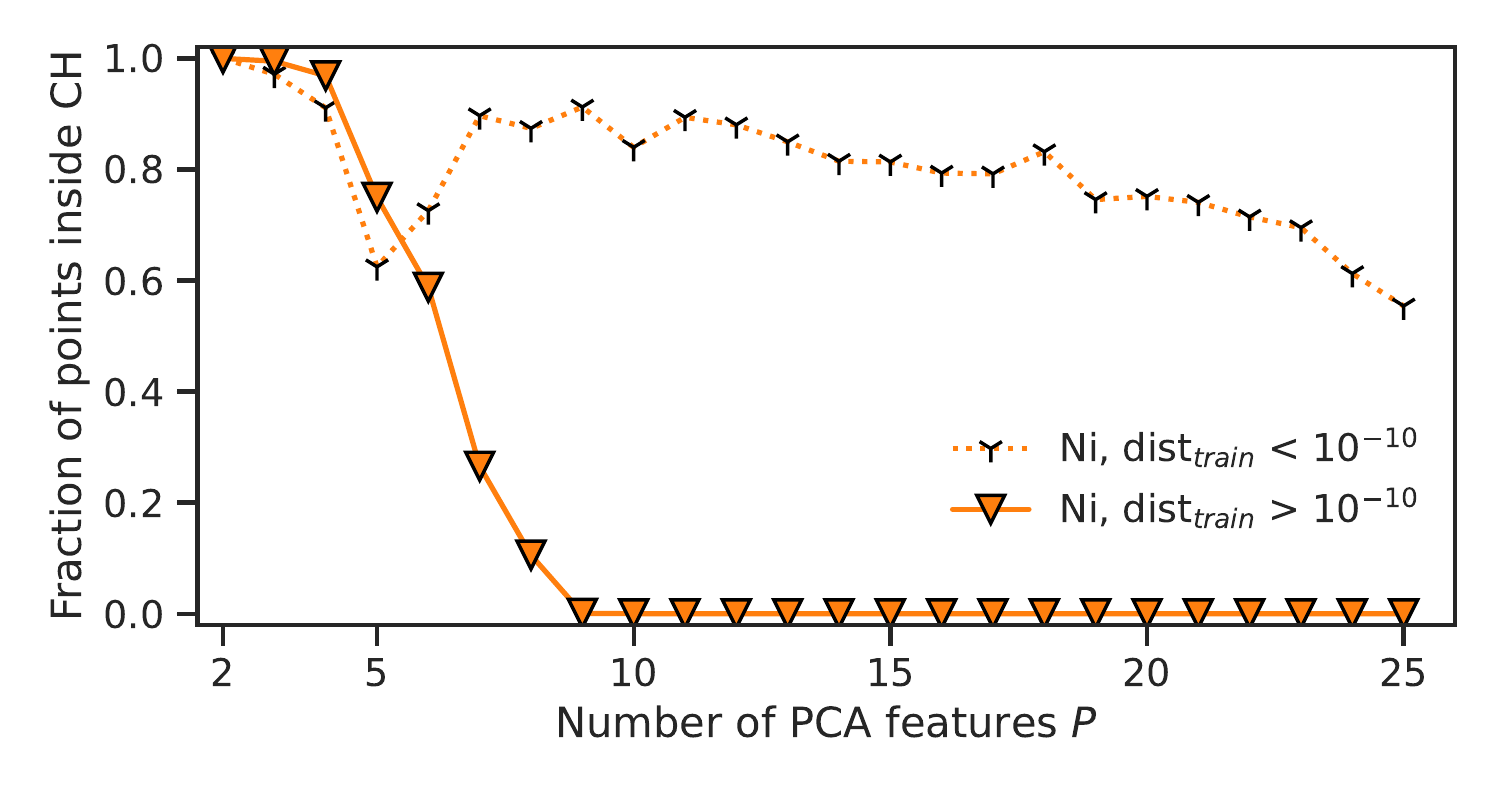}
    \caption{Fraction of atomic environments in the Ni test set which are contained in the N-dimensional CH enclosing the points in the training set, featurized according to the first $P$ PCA components of an ACE representation using  n$_{max}$ +l$_{max}$ =10.
    The dotted line refers to atomic environments in the test set that are distant less than $10^{-12}$ from at least one environment in the training set, the full line to the other environments in the test set.}
    \label{fig:ch_no_identical1}
\end{figure}
\begin{figure}[h!]
    \centering
    \includegraphics[width=8cm]{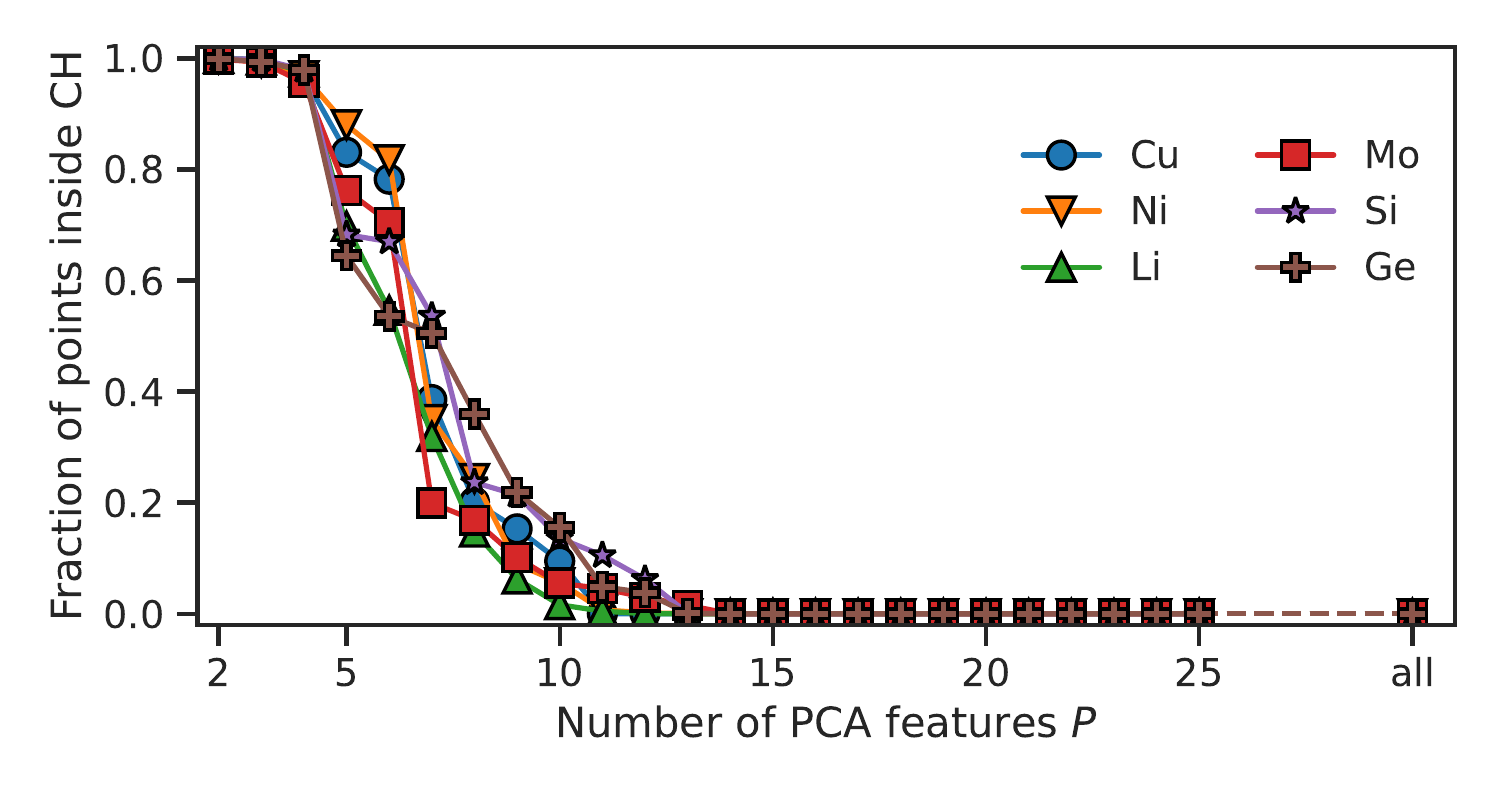}
    \caption{Fraction of atomic environments in the test set which are contained in the $P$-dimensional CH enclosing the points in the training set, featurized according to the first $P$ PCA components of an ACE representation using  n$_{max}$ +l$_{max}$ = 10.
    Atomic environments in the test set are less than 10$^{-12}$ distant from at least one environment in the training set have been excluded from the CH evaluation.}
    \label{fig:ch_no_identical2}
\end{figure}

\newpage

\subsection{Ridge Regression for Energy and Forces}

To train a machine learning model on forces and energies we build a 2D matrix $\mathbf{Y}$, whose elements $\mathbf{Y}_i$ correspond to atomic environments $S$ within a configuration $i$:
\begin{equation}
\mathbf{Y}_i = \left[ E_i, f^x_1, f^y_1, f^z_1, \dots, f^x_S, f^y_S, f^z_S \right].
\label{eq:force_energy_fitting}
\end{equation}
In the above $f^c_s$ labels the $c$-component of the force vector acting on the atom $s$ in the structure $i$.

The matrix of explanatory variables $\mathbf{Q}$ then takes the form of a 3D tensor with elements $\mathbf{Q}_i$ pertaining to structure $i$:
\begin{equation}
\mathbf{Q}_i = \left[ \mathbf{q}_i, -\dfrac{\partial \mathbf{q}_i}{\partial x_1}, -\dfrac{\partial \mathbf{q}_i}{\partial y_1}, -\dfrac{\partial \mathbf{q}_i}{\partial z_1}, \dots, -\dfrac{\partial \mathbf{q}_i}{\partial x_S}, -\dfrac{\partial \mathbf{q}_i}{\partial y_S}, -\dfrac{\partial \mathbf{q}_i}{\partial z_S} \right],
\label{eq:force_energy_descriptor}
\end{equation}
where $\mathbf{q}_i$ is the sum over all atoms $j$ in configuration $i$ of the atomic environment representation $\mathbf{q}(\rho_j)$.

When training a ridge regression machine learning model on energy and forces we translate the learning problem into the following closed formula:
\begin{equation}
    \mathbf{Y} = \mathbf{Q} ~ \mathbf{W}  + \mathbf{\epsilon},
\label{eq:ridge_form}
\end{equation}
where $\mathbf{Y}$ is the matrix of dependent variables , $\mathbf{Q}$ is the matrix of explanatory variables, $\mathbf{W}$ is the parameter matrix that weights $\mathbf{Q}$, and $\mathbf{\epsilon}$ is a vector of error terms which takes in consideration the possible presence of hidden variables which affect $\mathbf{Y}$ while not being encoded in $\mathbf{Q}$. 
For a given training set $\mathcal{D} = \{\mathbf{Y}_i, \mathbf{Q}_i \}~ i = 1, \dots, M$, one than finds the regression weights $\mathbf{W}$ analytically by solving:
\begin{equation}
\mathbf{W} = (\mathbf{Q}^T\mathbf{Q} + \gamma~ \mathbb{I})^{-1}\mathbf{Q}^T\mathbf{Y},
\label{eq:ridge_fitting}
\end{equation}
where $\gamma$ is the ridge parameter.

\newpage

\subsection{Regression weights for the Au$_{13}$ database}

\begin{figure}[h!]
    \centering
    \includegraphics[width=8cm]{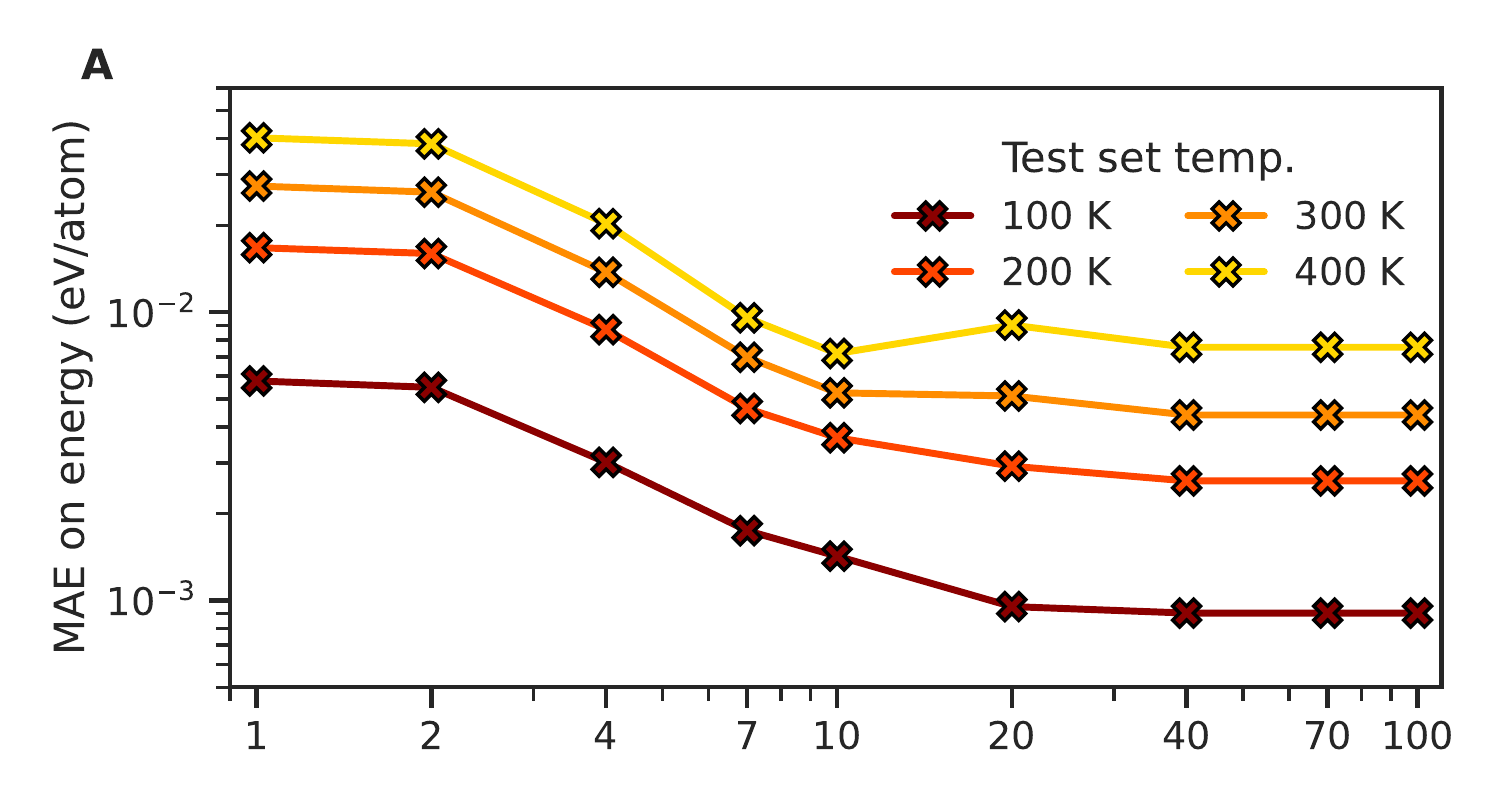}
    \includegraphics[width=8cm]{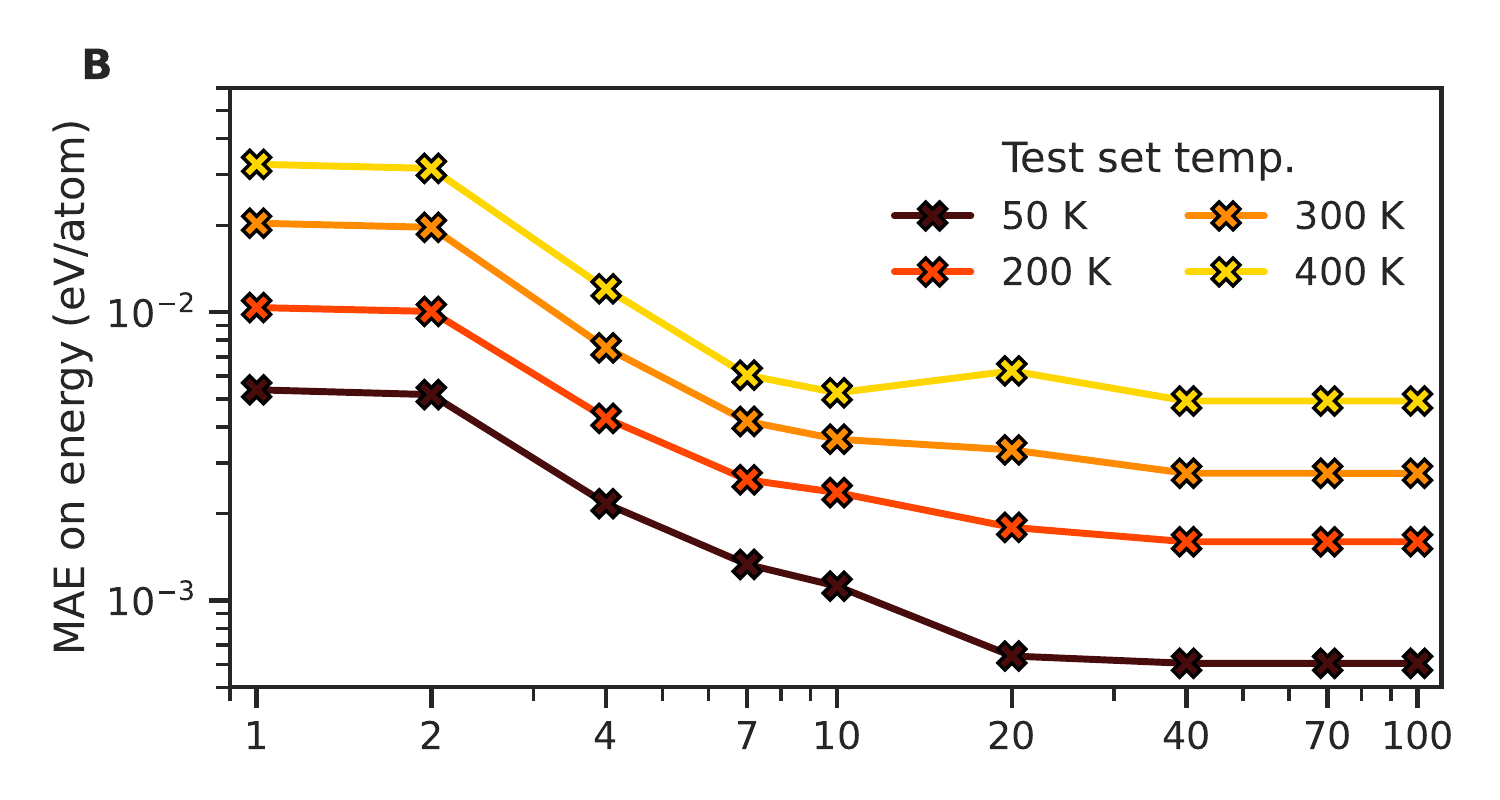}
    \includegraphics[width=8cm]{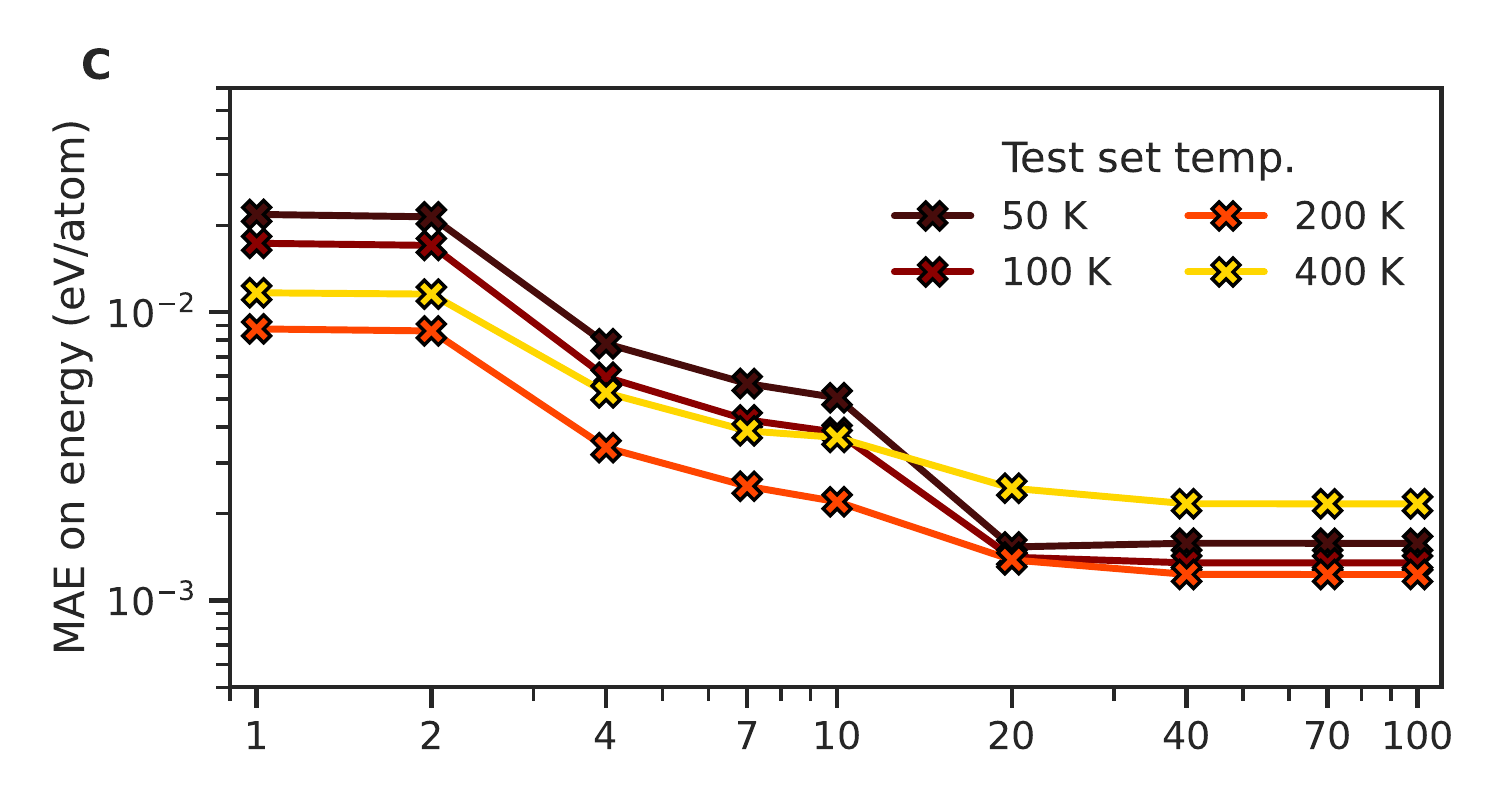}
    \includegraphics[width=8cm]{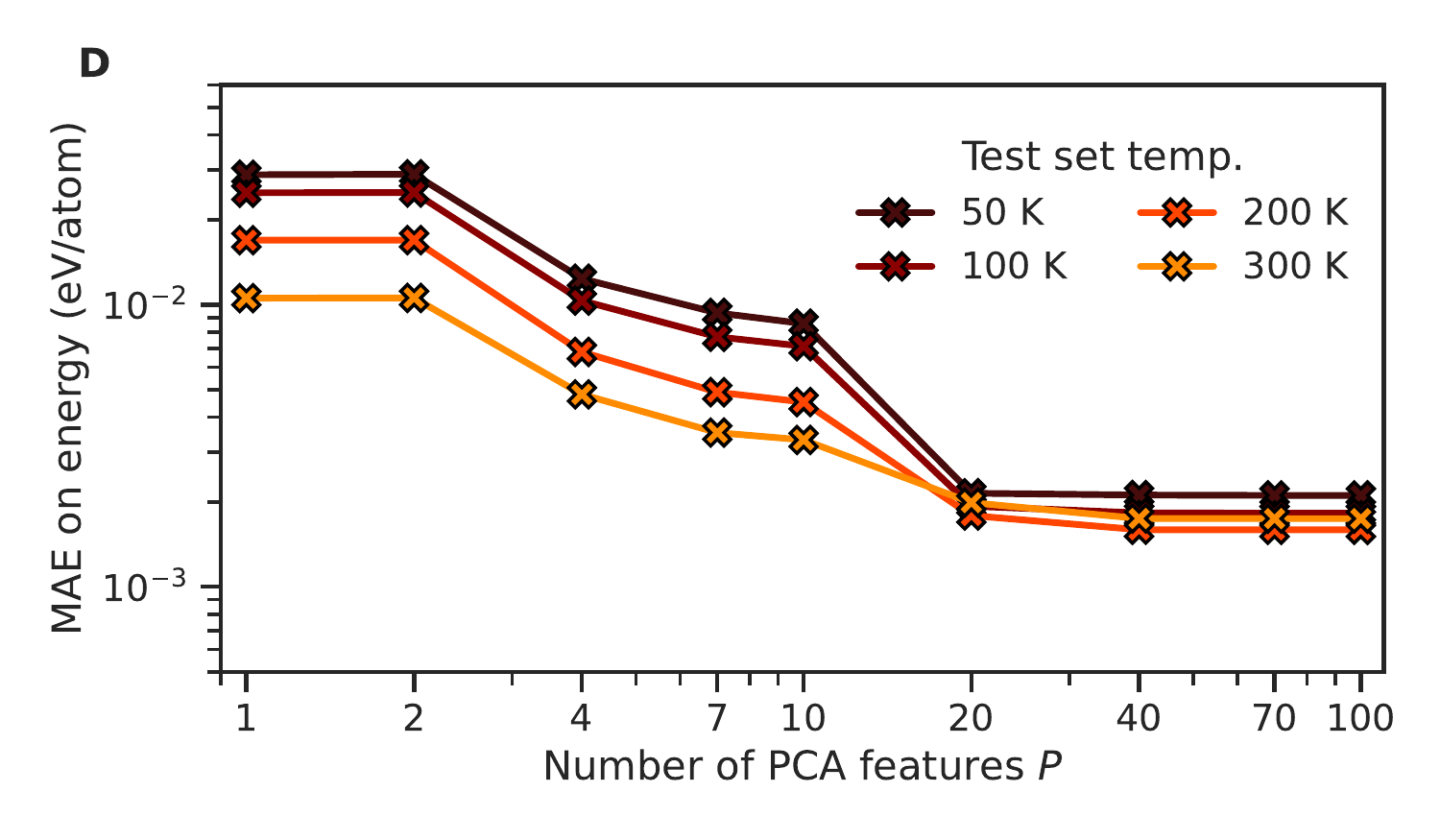}
    \caption{ Cumulative Mean Absolute Error on energy prediction when accounting for the contribution of the first $P$ PCA of the high-dimensional feature spaces deriving from an ACE representation for a ridge-regression potential trained on the Au$_{13}$ database at 50~K (Panel A),  100~K (Panel B),  300~K (Panel C),  400~K (Panel D).}
    \label{fig:cumulative-mae}
\end{figure}

\newpage

\subsection{Adaptive k-NN density estimation}

The density estimation method we used in this work consists of three steps.

First, we compute the intrinsic dimension of the dataset using the TwoNN estimator of Facco et al.~\cite{Facco2017}.
This approach is routed on the observation that, independently of the details of the density $\rho_i$ around each point $i$, the ratio $\mu_i = \frac{r_{i2}}{r_{i1}}$ of the second to first nearest neighbor distance follows a Pareto distribution which depends only on the intrinsic dimension $d$, and this allows to efficiently estimate $d$ by measuring the $\mu_i$'s.

Second, for each point $\mathbf{x}_i$ we compute the number of nearest neighbors $k_i$ over which the density can be considered approximately constant.
    This computation is performed through the likelihood ratio test proposed in Rodriguez et al.~\cite{Rodriguez2018}.
    In essence, the k-NN density of the point $i$ and that of its $(k+1)$th nearest neighbors are computed for increasing $k$ values. 
    If the two densities are too different, the test hypothesis of a constant density fails and the $k$th neighbor is selected as the optimal $k_i$ for point $i$.

Finally, the density $\rho$ of each point is estimated using via the simple k-NN equation
\begin{equation}
    \rho(\mathbf{x}_i) = \frac{k_i}{M V_i},
\end{equation}
In the above equation, $M$ is the number of points in the training set and $V_i$ is the spherical volume occupied by the $k_i$ nearest neighbor of $i$, computed as
\begin{equation}
    V_i = \omega_d r_{k_i}^d
\end{equation}
where $\omega_d$ volume of the $d$-sphere with unitary radius and $r_{k_i}$ is the distance between $i$ and its optimal $k_i$th nearest neighbor.

To evaluate the density of the training points on a point $\mathbf{x}^*$ \textit{belonging to the test set} we proceed as in ~\cite{Carli2021}.
We add $\mathbf{x}^*$ to the training set and perform the three steps described above, with the only difference that the final density on $\mathbf{x}^*$ is estimated as
\begin{equation}
    \rho(\mathbf{x}^*) = \frac{k^* - 1}{M V^*}.
\end{equation}
The code used in this work for the density estimations is available in open-source at ~\cite{DADApy}.

\newpage

\subsection{MAE and log density correlation}
\begin{figure}[h!]
    \centering
    \includegraphics[width=8.5cm]{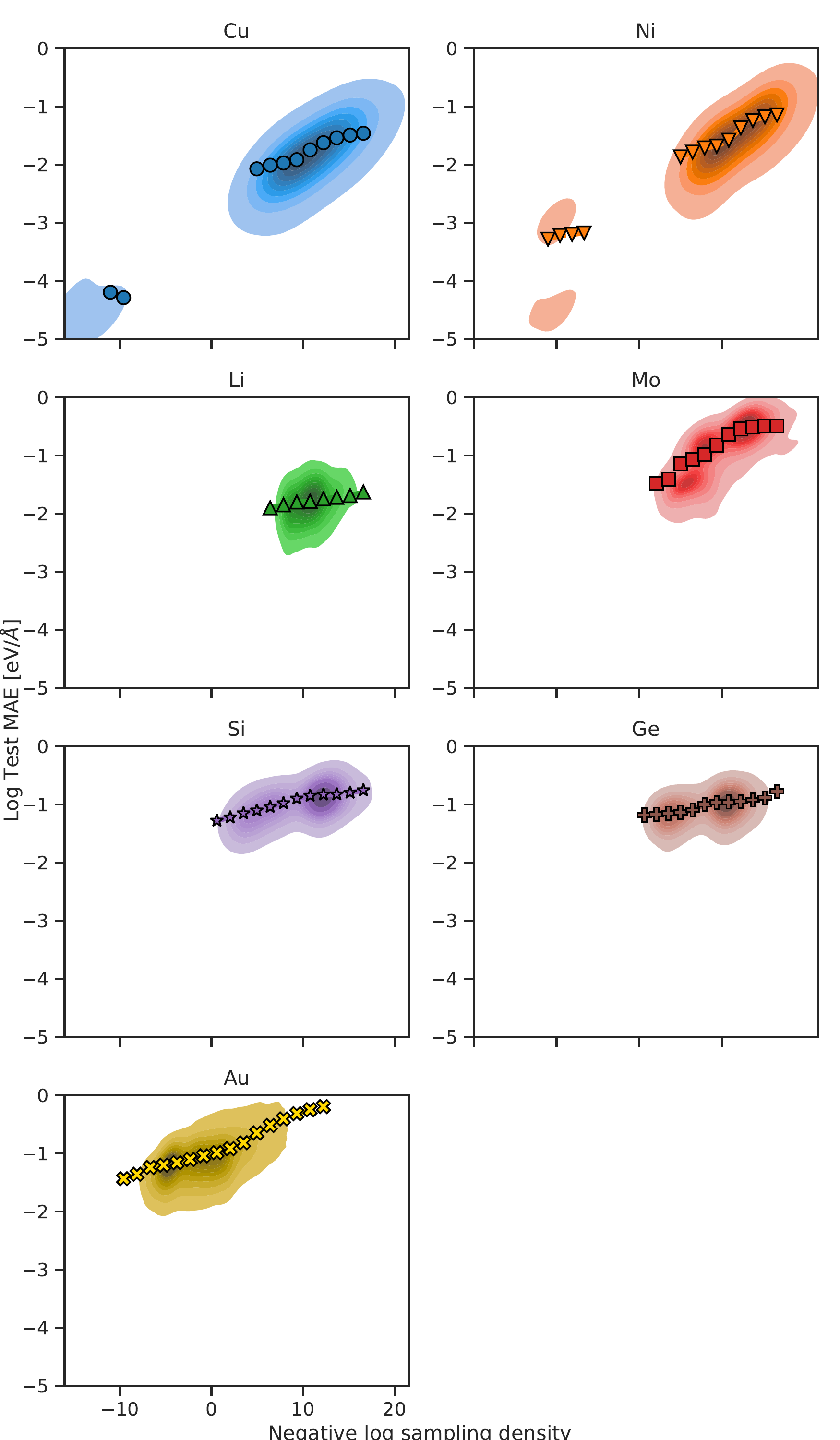}
    \caption{Two-dimensional density plot reporting the correlation between the test MAE incurred by ridge regression potentials in the validation set, and the negative probability density estimate for the training points in representation space, computed on test points.
    Darker colors indicate larger density of points.
    Points mark the average density bin, and are also reported in Figure 3 of the main text.
    The representation is an ACE one, with $N$ = 5 and n$_{max}$ + l$_{max}$ = 10.}
    \label{fig:rmse1}
\end{figure}

\begin{figure}[h!]
    \centering
    \includegraphics[width=8.5cm]{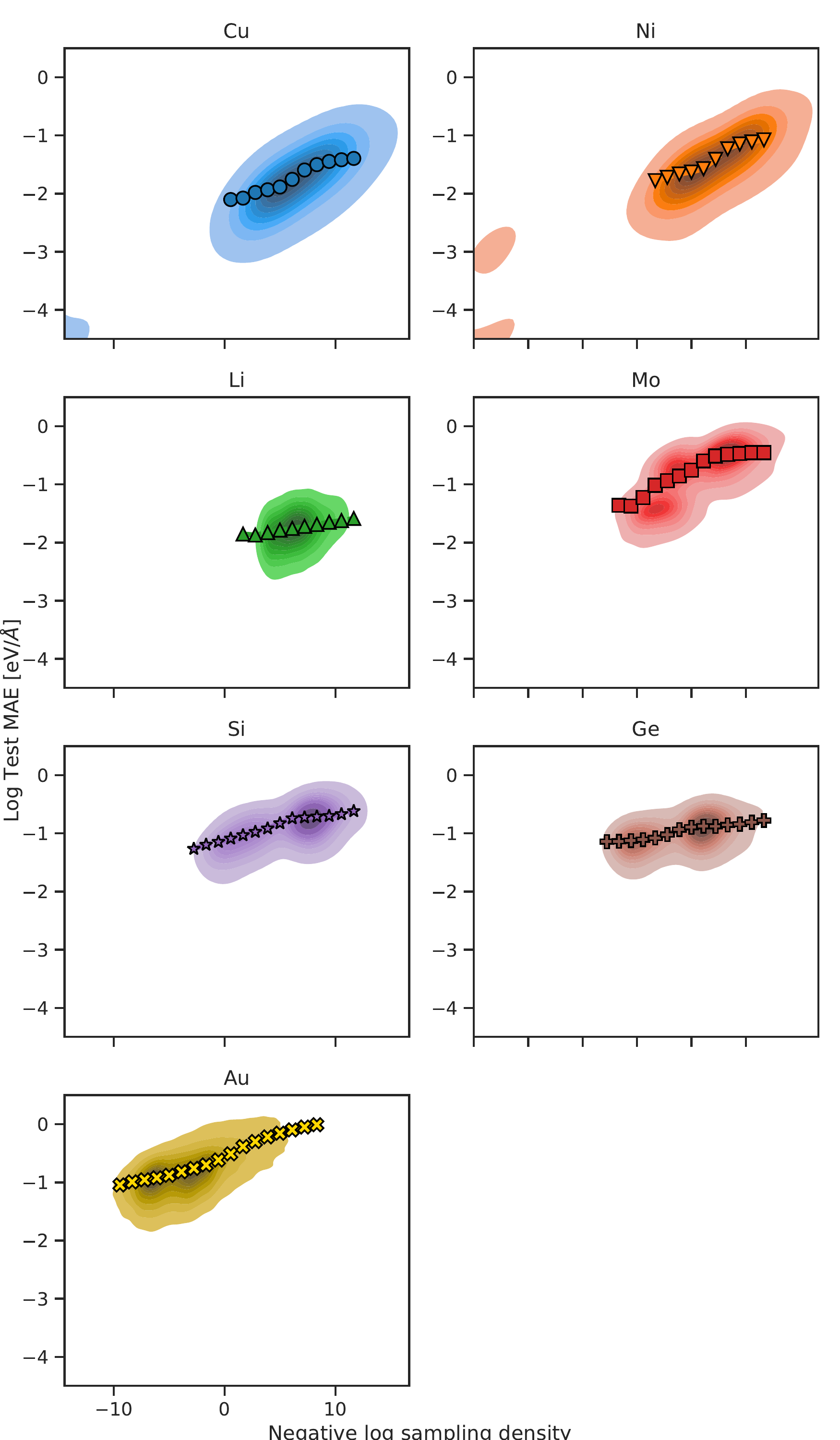}
    \caption{Two-dimensional density plot reporting the correlation between the test MAE incurred by ridge regression potentials in the validation set, and the negative probability density estimate for the training points in representation space, computed on test points.
    Darker colors indicate larger density of points.
    Points mark the average density binned over 20 equispaced bins, but only bins containing at least 1\% of the data are shown.
    The representation is an ACE one, with $N$ = 5 and n$_{max}$ + l$_{max}$ = 8.}
    \label{fig:rmse2}
\end{figure}

\begin{figure}[h!]
    \centering
    \includegraphics[width=8.5cm]{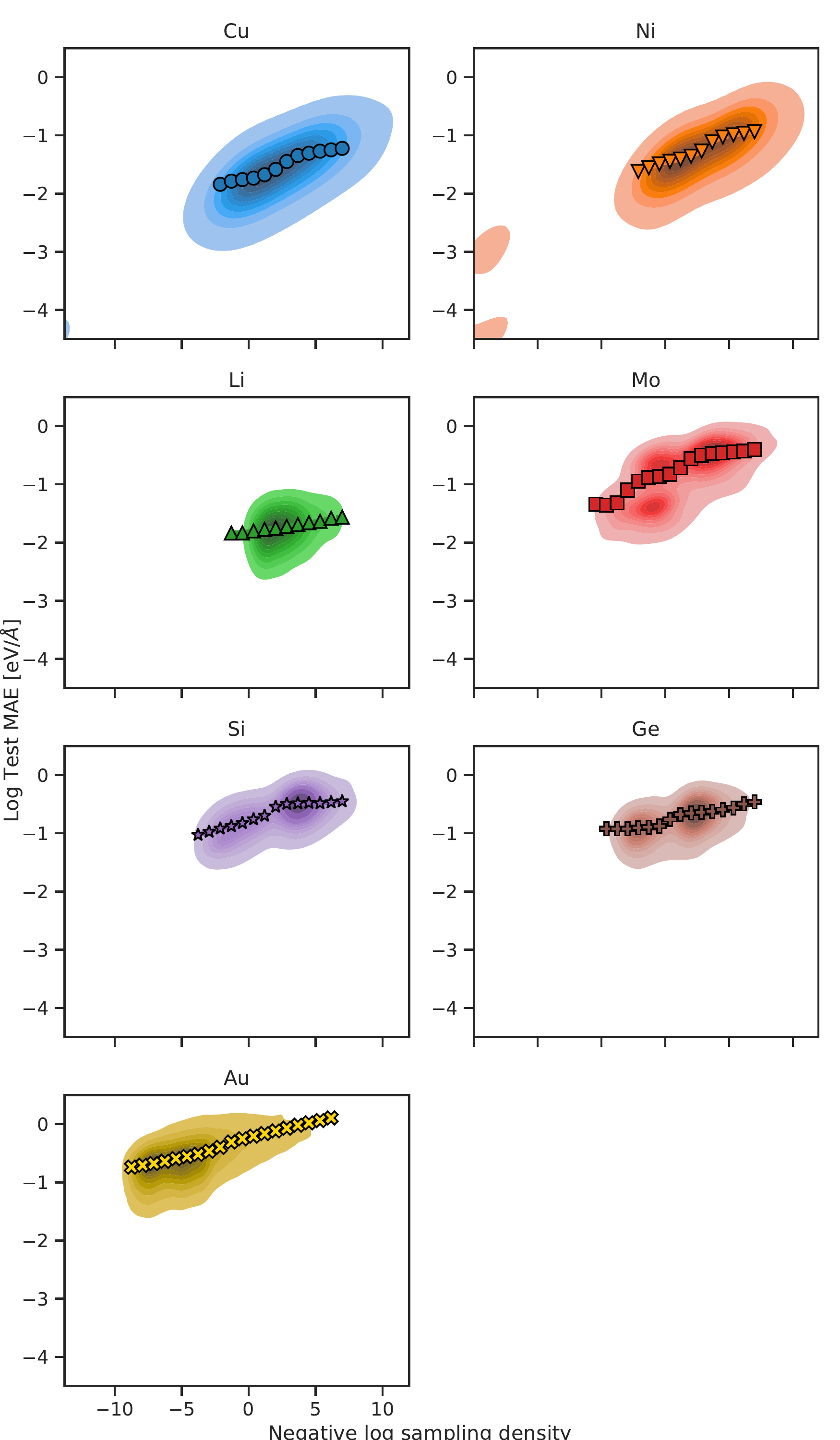}
    \caption{Two-dimensional density plot reporting the correlation between the test MAE incurred by ridge regression potentials in the validation set, and the negative probability density estimate for the training points in representation space, computed on test points.
    Darker colors indicate larger density of points.
    Points mark the average density binned over 20 equispaced bins, but only bins containing at least 1\% of the data are shown.
    The representations is an ACE one, with $N$ = 5 and n$_{max}$ + l$_{max}$ = 6.}
    \label{fig:rmse3}
\end{figure}

Figures \ref{fig:rmse1}, \ref{fig:rmse2}, and \ref{fig:rmse3} report the two-dimensional density plot for the \citet{zuo2020performance} and the Au$_{13}$ datasets,  highlighting the correlation between the test MAE incurred by ridge regression potentials in the validation set, and the negative probability density estimate for the training points in representation space, computed on test points.
We do so for the case of ACE descriptors with N=5 and and n$_{max}$ + l$_{max}$ = 10 (Figure \ref{fig:rmse1}), 8 (Figure \ref{fig:rmse2}), and 6 (Figure \ref{fig:rmse3}).

\newpage

\newpage

\newpage

\subsection{Comparison of adaptive k-nn with other error estimation methods}

We here present a comparison of the metric based on the density-estimation method w.r.t. other ones that have been also used as a proxy for the predictiveness of a model.
In particular, we look at  the negative log density computed using kernel density estimation (KDE) \cite{davis2011remarks}
 (subsection K1), the logarithm of the minimal distance from the training set \cite{Imbalzano2018} (subsection K2), and the logarithm of the standard deviation of predictions yielded by a committee of ridge regression models \cite{Imbalzano2021} (subsection K3).
We compute the above properties for each test point.

In order to compare quantitatively how well a given metric estimates the error incurred by a potential on a test local atomic environment, we look at the spread of MAEs as a function of each metric's value.
We expect an ideal metric to have very low MAE's standard deviation across all its values.
This would indeed indicate that such metric is tightly correlated with the MAE, and that a function can be constructed to predict the MAE with high confidence.

To compare quantities estimated according to each metric on the same footing, we normalize their values so to their distribution has zero mean and unitary variance.
We then bin each metric into 31 equally spaced bins between -3 and 3, of width 0.4, and compute the ratio between the standard deviation and the average of the MAE incurred on the test set, for all local atomic environments lying inside each bin.

\begin{figure}[h!]
    \centering
    \includegraphics[width=8.5cm]{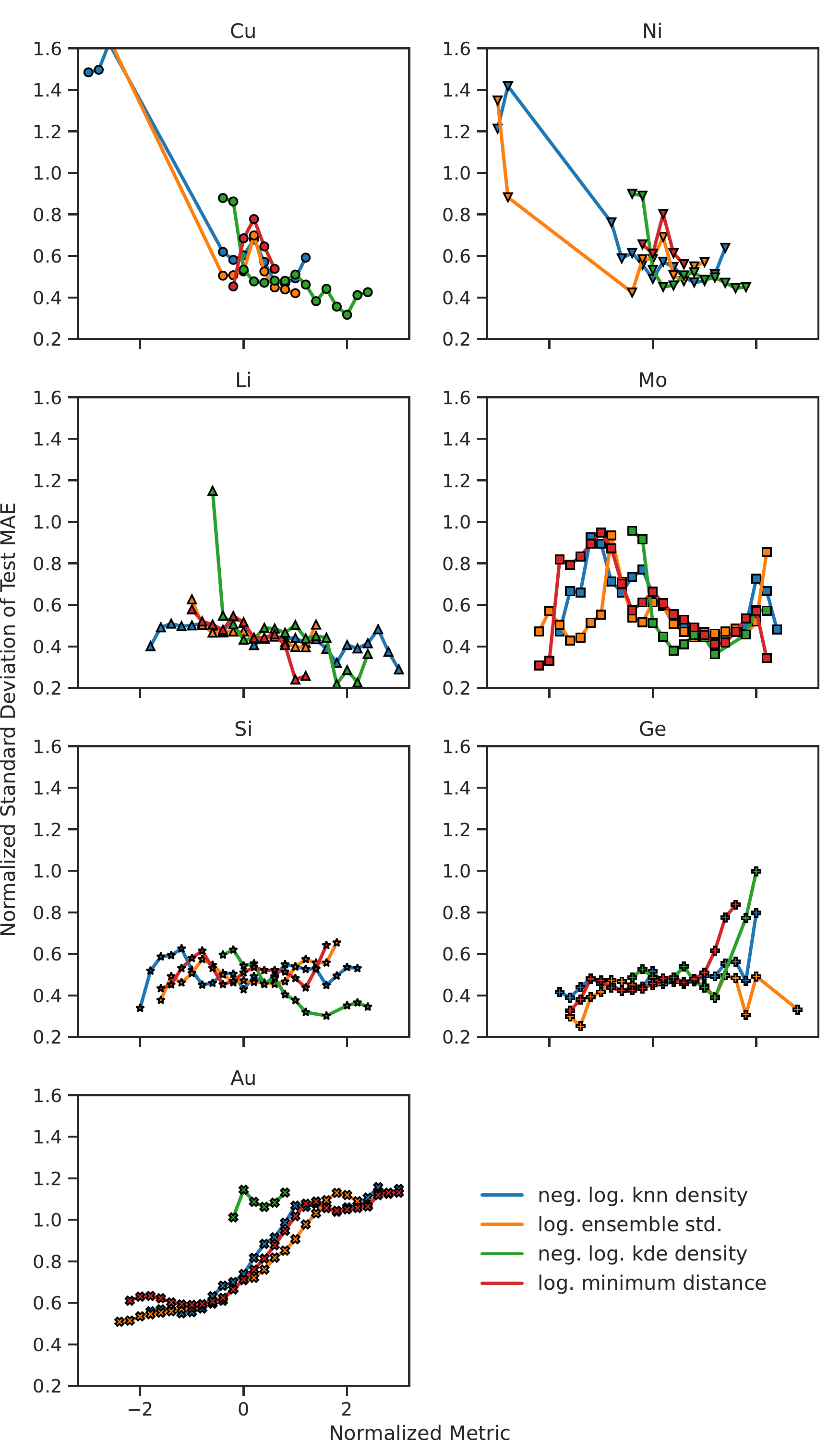}
    \caption{Correlation between the relative spread (standard deviation of MAE divided by average MAE) of the distribution of MAEs and four metrics for the predictive variance estimate.
    The four metrics are: the negative log sampling density evaluated using adaptive k-nn (blue), the logarithm of the standard deviation for force predictions incurred by an ensemble of 5 ridge regression potentials (orange), the negative log sampling density evaluated via KDE (green), and the logarithm of the minimum distance form the training set in representation space (red).
    Points mark the standard deviation of the MAE binned over 30 equispaced bins, but only bins containing at least 0.5\% of the data are shown.
    The representations is an ACE one, with $N$ = 5 and n$_{max}$ + l$_{max}$ = 10.}
    \label{fig:metric_comparison}
\end{figure}

In Figure \ref{fig:metric_comparison}, we plot the correlation existing between the aforementioned normalized metrics, as well as the negative log density we employ in the main text w.r.t., and the standard deviation of the force MAE incurred on the test-set by a ridge regression potential employing an ACE representation with $N$ = 5 and n$_{max}$ + l$_{max}$ = 10.
We observe that the performance of all metrics, except KDE, is comparable.
This is especially evident for the case of Au$_{13}$, where the negative log of training set density evaluated via adaptive k-nn, the log of the minimum distance from the training set, and the log of the standard deviation incurred by an ensemble of potentials perform almost identically.

More in general, the analysis displayed in Figure \ref{fig:metric_comparison} indicates that the three aforementioned metrics can be also employed to estimate the MAE incurred by a ridge regression potential, as they tend to predict the force MAE on test points with a standard deviation that is approximately half of its average value (y values around 0.5 in Figure \ref{fig:metric_comparison}).

\newpage

\newpage


\subsubsection{Comparison with kernel density estimation}

\begin{figure}[h!]
    \centering
    \includegraphics[width=7.5cm]{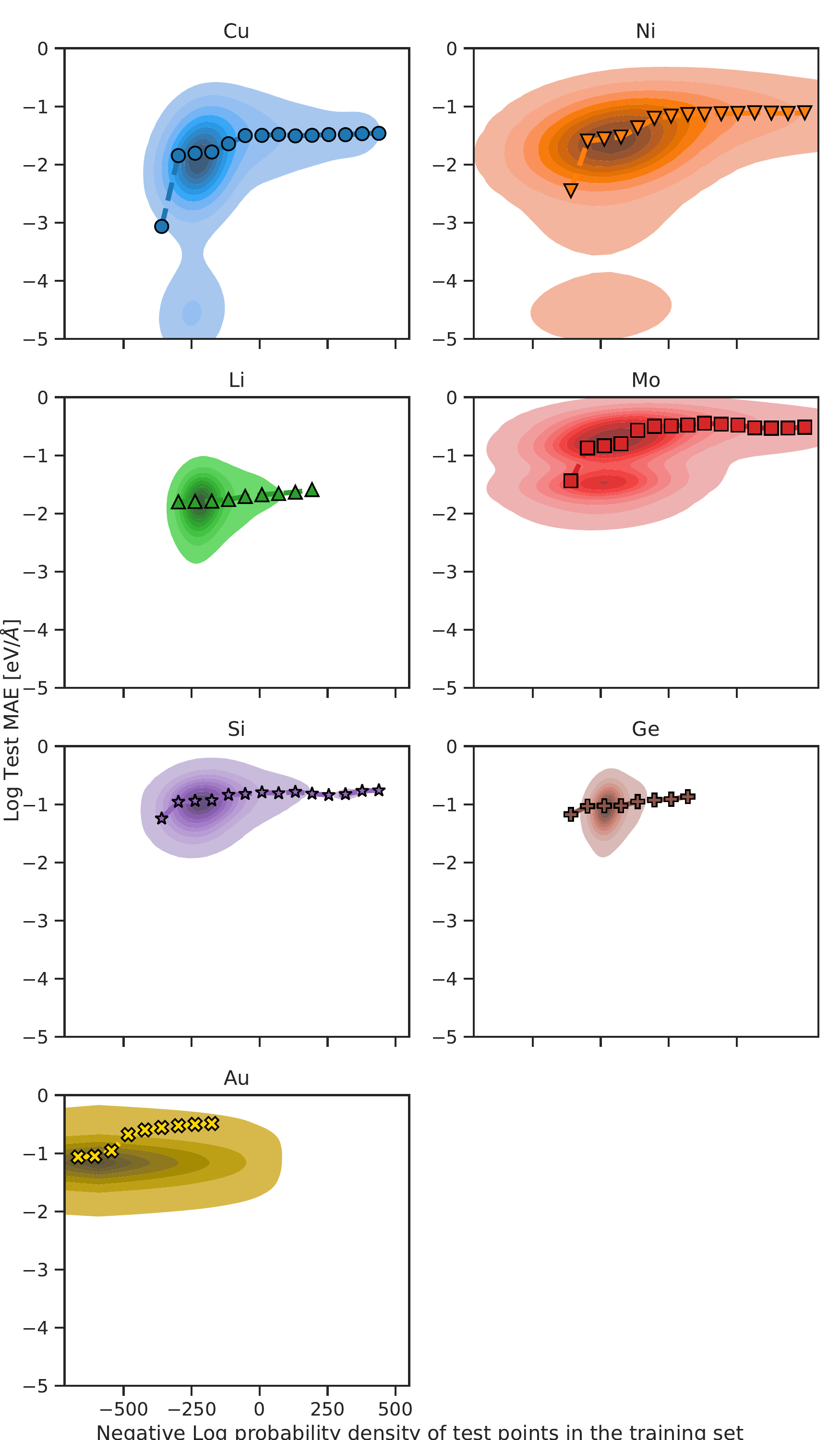}
    \caption{Two-dimensional density plot reporting the correlation between the test MAE incurred by ridge regression potentials in the validation set, and the probability density estimate for the training points in representation space, computed on test points.
    Densities are calculated with a vanilla implementation of the k-NN density estimator, which disregards the intrinsic dimensionality where the data lie.
    Darker colors indicate larger density of points.
    Points mark the average density binned over 20 equispaced bins, but only bins containing at least 1\% of the data are shown.
    The representation is an ACE one, with $N$ = 5 and n$_{max}$ + l$_{max}$ = 10.}
    \label{fig:kde_density}
\end{figure}

We compare the performance of the adaptive k-nn density estimation method we employ with kernel density estimation (KDE) \cite{davis2011remarks}.
To this end, we employ a Gaussian kernel and look at bandwidth values ranging from 10$^{-6}$ to 10$^2$.
In Figure \ref{fig:kde_density} we showcase that a weak correlation between the negative log density of test set points evaluated via KDE and the error incurred by the ML model arises.
We observe that disregarding the intrinsic dimensionality of the manifold where data points lie results in densities whose magnitude is astonishingly small, in agreement with previous report in the literature \cite{Martiniani2016}.
This result highlights the advantage of employing a density estimation method that calculates and uses the intrinsic dimensionality of the data manifold.

\newpage

\newpage

\subsubsection{Comparison with minimal distance from training set}

\begin{figure}[h!]
    \centering
    \includegraphics[width=7.5cm]{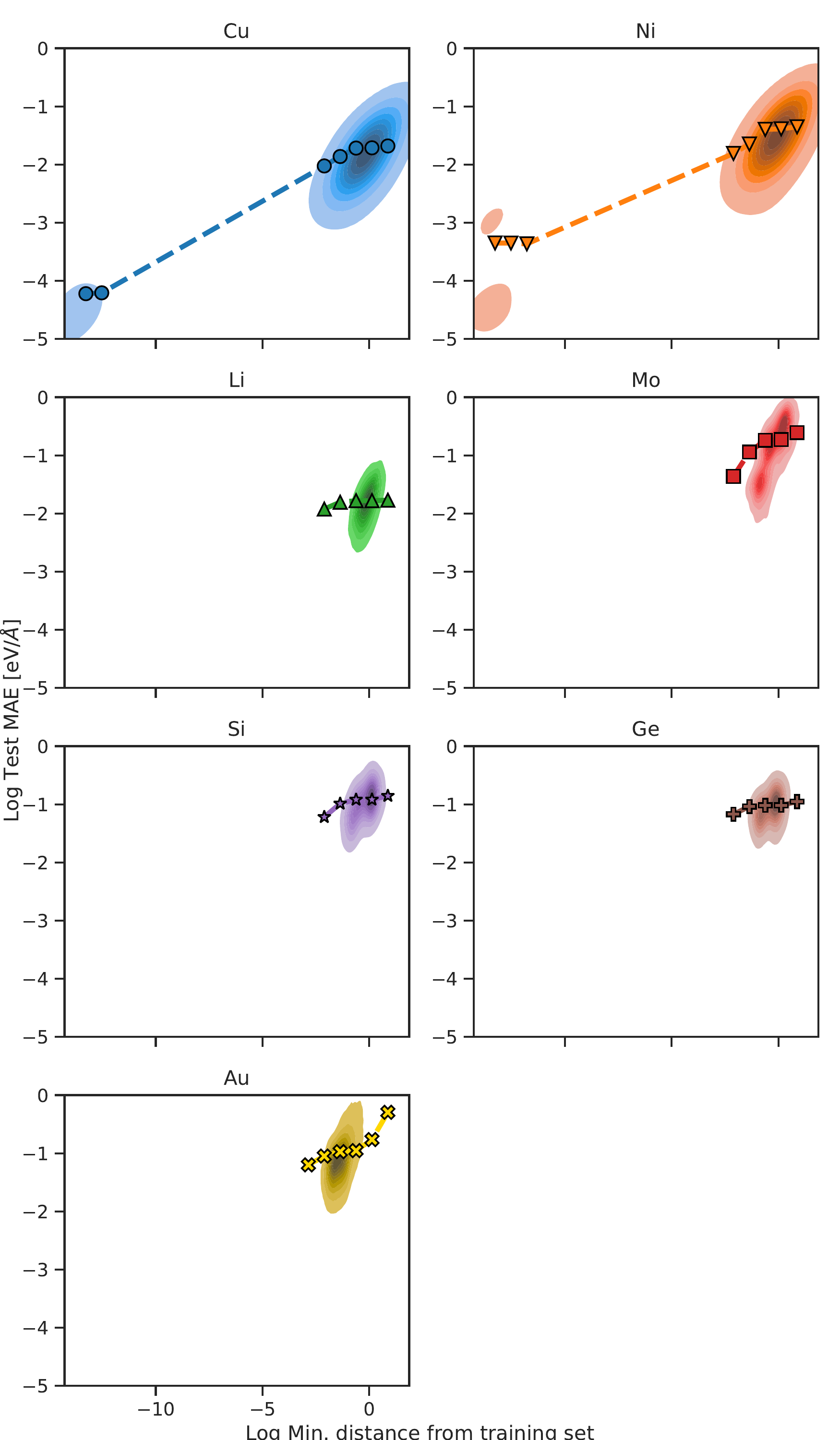}
    \caption{Two-dimensional density plot reporting the correlation between the test MAE incurred by ridge regression potentials in the validation set, and the  logarithm of the minimum Euclidean distance of test points from the training set in the descriptor space.
    Darker colors indicate larger density of points.
    Points mark the average density binned over 20 equispaced bins, but only bins containing at least 1\% of the data are shown.
    The representation is an ACE one, with $N$ = 5 and n$_{max}$ + l$_{max}$ = 10.}
    \label{fig:hausdorrf_density}
\end{figure}

Any density estimate method for a set of points will yield a  probability density that is, in some way, a function of the distance from the points in the set.
The adaptive k-nn density estimation method we employ is no exception, and here we analyze how such method behaves similarly to the negative logarithm of the minimum Euclidean distance from the training set.
In Figure \ref{fig:hausdorrf_density}, we report the two-dimensional plot showcasing the existing correlation between the MAE on forces incurred by a ridge regression potential for each atom in the test set, and the minimal distance of the descriptor associated to such atom and its local environment to each other descriptor in the training set.

\newpage

\newpage

\newpage

\subsubsection{Comparison with standard deviation of committee ensemble}

\begin{figure}[h!]
    \centering
    \includegraphics[width=8.5cm]{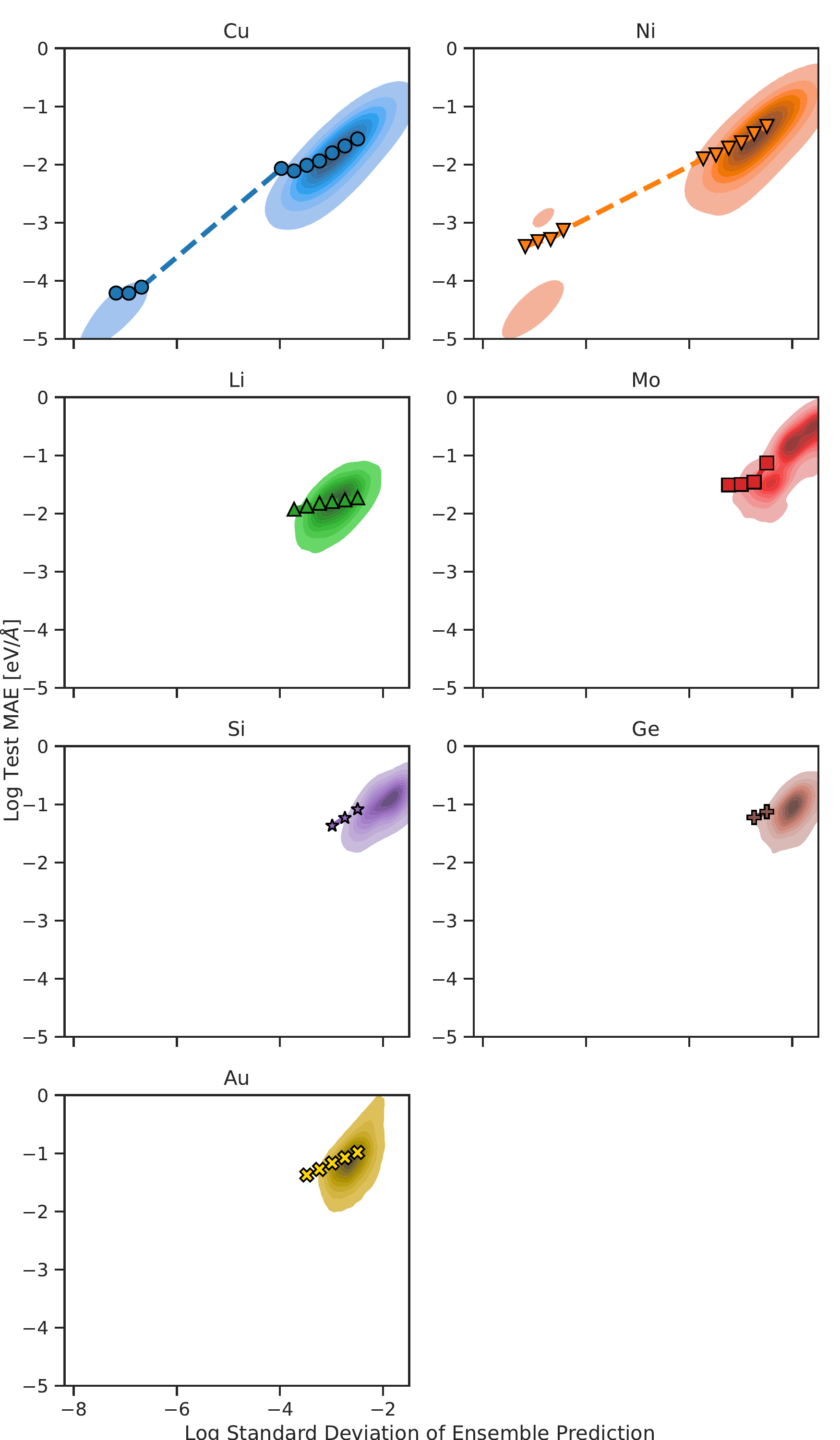}
    \caption{Two-dimensional density plot reporting the correlation between the test MAE incurred by ridge regression potentials in the validation set, and the log standard deviation for force predictions on the test sets by a committee of five ridge regression potentials.
    Darker colors indicate larger density of points.
    Points mark the average density binned over 20 equispaced bins, but only bins containing at least 1\% of the data are shown.
    The representation is an ACE one, with $N$ = 5 and n$_{max}$ + l$_{max}$ = 10.}
    \label{fig:std_mae}
\end{figure}

We look at the correlation between the log density of points in the test set computed using the adaptive k-nn density estimation fitted to the training set, and the standard deviation of an ensemble of linear regression potentials on the test set points.
To do so, we calculate the standard deviation on the force predictions yielded by 5 ridge regression potentials trained upon the ACE representations with N=5 and n$_{max}$ + l$_{max}$ = 10.
These 5 ridge regression potentials have each been trained on 80\% of the training data, chosen at random.
In Figure \ref{fig:std_mae}, we report the two-dimensional density plot for the logarithm of the standard deviation for the test set force predictions made by such set of 5 ridge regression potentials, and the logarithm of the force error incurred by another ridge regression potential, trained on the whole training set, on a test set.
We observe a good correlation, as is expected by an ensemble method, and as previously reported by \citet{Imbalzano2018} for the case of artificial neural network potentials.
We remark that such predictive error estimate algorithm is method-dependent, and computationally intensive (i.e., it demands to train multiple ML models) .\\

\newpage

\end{document}